\documentclass[a4paper,british,english]{IEEEtran}
\usepackage[T1]{fontenc}
\usepackage[latin9]{inputenc}
\usepackage{mathrsfs}
\usepackage{amsmath}
\usepackage{amssymb}

\makeatletter

\pdfpageheight\paperheight
\pdfpagewidth\paperwidth

\usepackage[printonlyused,withpage]{acronym}

\usepackage{ifthen}
\providecommand{\nomenclature}{}
\renewcommand{\nomenclature}[2]{\ifthenelse{\equal{#2}{f}}{\acf{#1}}{\ifthenelse{\equal{#2}{s}}{\acs{#1}}{\ifthenelse{\equal{#2}{l}}{\acl{#1}}{\ifthenelse{\equal{#2}{p}}{\acp{#1}}{\ac{#1}}}}}}

\usepackage{enumitem}
\usepackage{framed}

\usepackage{graphicx}

\usepackage{tikz}
\usetikzlibrary{matrix,shapes,arrows,positioning,chains}

\usetikzlibrary{ arrows,  shapes.misc,  shapes.arrows,  chains,  matrix,  positioning,  scopes,  decorations.pathmorphing,  shadows, calc}
\tikzset{
nonterminal/.style={rectangle, minimum size=6mm, very thick, draw=red!50!black!20, top color=white, bottom color=red!50!black!50, font=\itshape },
terminal/.style={rounded rectangle,  minimum size=6mm, very thick,draw=black!50, top color=white, bottom color=black!20, font=\ttfamily},
junction/.style={circle, draw }}
\pgfsetlinewidth{.4mm}

\usetikzlibrary{decorations.pathmorphing} 
\usetikzlibrary{matrix} 
\usetikzlibrary{arrows} 
\usetikzlibrary{calc} 

\tikzstyle{block} = [draw,rectangle,thick,minimum height=2em,minimum width=2em]
\tikzstyle{sum} = [draw,circle,inner sep=0mm,minimum size=2mm]
\tikzstyle{connector} = [->,thick]
\tikzstyle{line} = [thick]
\tikzstyle{branch} = [circle,inner sep=0pt,minimum size=1mm,fill=black,draw=black]
\tikzstyle{branch2} = [circle,inner sep=0pt,minimum size=0mm,fill=black,draw=black]
\tikzstyle{guide} = []
\tikzstyle{snakeline} = [connector, decorate, decoration={pre length=0.2cm,
                         post length=0.2cm, snake, amplitude=.4mm,
                         segment length=2mm},thick, magenta, ->]

\tikzset{
    state/.style={
           rectangle,
           rounded corners,
           draw=black, very thick,
           minimum height=1em,
           inner sep=1pt,
           text centered,
           },
}

\usetikzlibrary{calc,backgrounds}
\usepackage{verbatim}

\usetikzlibrary{positioning}
\usetikzlibrary{decorations.text}
\usetikzlibrary{decorations.pathmorphing}

\usepackage{bm}

\usepackage{ctable}

\usetikzlibrary{fit}					
\usetikzlibrary{backgrounds}	

\usepackage{lettrine}

\makeatother

\usepackage{babel}

\usepackage{fancyhdr}
\fancypagestyle{firstpage}{
	\fancyhf{}
	\fancyhead[L]{{\small SUBMITTED TO IEEE/ACM TRANSACTIONS ON AUDIO, SPEECH AND LANGUAGE PROCESSING}}
    
    \fancyhead[R]{\thepage}
	
	\fancyfoot[C]{}
}

\begin{document}

\title{P\MakeLowercase{hase}-A\MakeLowercase{ware} S\MakeLowercase{ingle}-C\MakeLowercase{hannel}
S\MakeLowercase{peech} E\MakeLowercase{nhancement} \MakeLowercase{with}
M\MakeLowercase{odulation}-D\MakeLowercase{omain} K\MakeLowercase{alman}
F\MakeLowercase{iltering} }

\author{Nikolaos Dionelis, Student Member, IEEE, Mike Brookes, Member, IEEE}
\maketitle

\thispagestyle{firstpage}

\begin{abstract}
We present a single-channel phase-sensitive speech enhancement algorithm
that is based on modulation-domain Kalman filtering and on tracking
the speech phase using circular statistics. With Kalman filtering,
using that speech and noise are additive in the complex STFT domain,
the algorithm tracks the speech log-spectrum, the noise log-spectrum
and the speech phase. Joint amplitude and phase estimation of speech
is performed. Given the noisy speech signal, conventional algorithms
use the noisy phase for signal reconstruction approximating the speech
phase with the noisy phase. In the proposed Kalman filtering algorithm,
the speech phase posterior is used to create an enhanced speech phase
spectrum for signal reconstruction. The Kalman filter prediction models
the temporal/inter-frame correlation of the speech and noise log-spectra
and of the speech phase, while the Kalman filter update models their
nonlinear relations. With the proposed algorithm, speech is tracked
and estimated both in the log-spectral and spectral phase domains.
The algorithm is evaluated in terms of speech quality and different
algorithm configurations, dependent on the signal model, are compared
in different noise types. Experimental results show that the proposed
algorithm outperforms traditional enhancement algorithms over a range
of SNRs for various noise types. 
\end{abstract}

\begin{IEEEkeywords}
\textmd{{Speech enhancement; single-channel; speech phase} }
\end{IEEEkeywords}

\IEEEpeerreviewmaketitle{
\begin{table}[b]

\IEEEpeerreviewmaketitle{\quad N. Dionelis and M. Brookes are with the Department of Electrical and Electronic Engineering, Imperial College London, London SW7 2AZ, U.K. (e-mail: nikolaos.dionelis11@imperial.ac.uk; mike.brookes@imperial.ac.uk). }
\end{table}
}

\section{Introduction }

\label{sec:intro}\lettrine[lines=2]{S}{peech} enhancement in non-stationary
noise environments is a challenging research area. The modulation
domain is an often-used representation in models of the human auditory
system; in speech enhancement, the modulation domain models the temporal/inter-frame
correlation of frames rather than treating each frame independently
\cite{a203} \cite{a233}. Enhancement algorithms can benefit from
including a model of the inter-frame correlation of speech and a number
of authors have found that the performance of a speech enhancer can
be improved by using a speech model that imposes temporal structure
\cite{a47}, \cite{a185}, \cite{a222}. Temporal inter-frame speech
correlation modelling can be performed with a Kalman filter (KF) with
a state of low dimension, as in \cite{a203} and \cite{a126}. The
enhancement algorithms in \cite{a123} track the time evolution of
the clean Short Time Fourier Transform (STFT) amplitude domain coefficients
in every frequency bin. In \cite{a174}, speech inter-frame correlation
is modeled. Considering KF algorithms, many papers, such as \cite{a132}
\cite{a133} and \cite{a129}, utilise the nonlinear observation model
relating clean and noisy speech in the log-spectral domain. 

An overview of traditional statistical-based algorithms is given in
\cite{a184}. Whereas traditional statistical-based algorithms treat
each time-frame independently, an alternative approach performs filtering
in the modulation domain to model the time correlation of speech.
Inter-frame speech correlation modelling can help enhancement algorithms
achieve more noise reduction and less speech distortion \cite{a247}.
Modulation-domain Kalman filtering refers to sequentially updating
the statistics of speech utilising temporal constraints on the estimation
of the speech spectrum with a KF prediction step \cite{a126}, \cite{a123},
\cite{a225}. In \cite{a37} and \cite{a13}, KF tracking of speech
(and noise in \cite{a192}) is presented. A number of authors perform
speech Kalman filtering in the amplitude spectral domain assuming
additivity of speech and noise in the same spectral domain \cite{a203},
\cite{a204}. 

Existing enhancement KF algorithms that work in the time-frequency
domain differ in their choice of the KF state, the KF prediction and
the KF update. The KF state can be in the speech amplitude spectral
domain \cite{a203} \cite{a204}, the power spectral domain or the
log-spectral domain \cite{a225} \cite{a236}. Speech spectra are
well modelled by Gaussians in the log-spectral domain (and not so
well in other domains), mean squared errors in the log-spectral domain
are a good measure to use for perceptual speech quality and the non-nonnegative
log-spectral domain is most suitable for infinite-support Gaussian
modelling. 

Regarding the KF prediction, AR modelling with or without the AR mean
can be performed based on the autocorrelation method, which uses windowing
\cite{a144}, or on the covariance method allowing or not allowing
unstable AR poles. A KF with non-time-varying parameters requires
the use of stable AR poles in the KF prediction, while a KF with time-varying
parameters does not require the use of stable AR poles. Despite the
observation that a time-varying KF does not require stable AR poles,
the stability of the KF loop is still important. 

The KF update is affected by the signal model that is used for the
addition of speech and noise \cite{a109}. If speech and noise are
independent, then they add in the complex STFT domain \cite{a111}
\cite{a226}; it may however be analytically simpler to assume that
speech and noise add either in the power spectral domain or the amplitude
spectral domain \cite{a203} \cite{a204}. The aforementioned alternative
possible ways are related to the phase factor, which is the cosine
of the phase difference between speech and noise \cite{a111} \cite{a181}.
Assuming speech and noise additivity in the power domain is equivalent
to assuming that the phase factor is zero. Assuming speech and noise
additivity in the amplitude domain is equivalent to assuming that
the phase factor is equal to unity. The Kalman filtering algorithms
in \cite{a203} and \cite{a204} assume that speech and noise are
additive in the amplitude spectral domain, which results in noise
oversubtraction in the region of $0$ dB SNR that may sometimes be
perceptually good \cite{a184}. 

Noise tracking using a KF can be beneficial for enhancement \cite{a243},
\cite{a192}. Noise tracking is performed in \cite{a243} and subsequently
in \cite{a241}. In the KF update step, the correlation between speech
and noise can be estimated \cite{a225}, \cite{a236}. 

Enhancement algorithms can benefit either from modelling the phase
factor \cite{a225} or from estimating the speech phase \cite{a217}.
The two related but distinct issues here are: (a) modelling the phase
factor allows a better estimate of the speech amplitude, and (b) modelling
the phase itself allows a better estimate of the speech phase. Both
(a) and (b) can potentially improve the performance of the speech
enhancer. Amongst other technical papers, the phase factor is considered
in \cite{a111} and \cite{a238}. On the contrary, in \cite{a129},
the phase factor is assumed to be zero. Estimating the speech phase
and not estimating the noise phase does not affect the phase factor
distribution; if the noise phase is uniformly distributed, then the
difference of the speech phase and the noise phase is also uniformly
distributed. 

Recent research has shown the importance of estimating the speech
phase \cite{a217} \cite{a11}: approximating the speech phase with
the noisy phase can hinder the performance of enhancement algorithms.
In recent years, the estimation of the speech phase is a widespread
problem in speech processing \cite{a217}. The estimation of the speech
phase as a circular quantity using circular probability distributions
is a less widespread problem. Circular statistics take phase periodicity
explicitly into account so that speech phase tracking does not fail
when the discontinuity between $- \pi$ and $\pi$ is reached \cite{a215}. 

\begin{figure}[!t]
\centering \includegraphics[width=0.61\columnwidth]{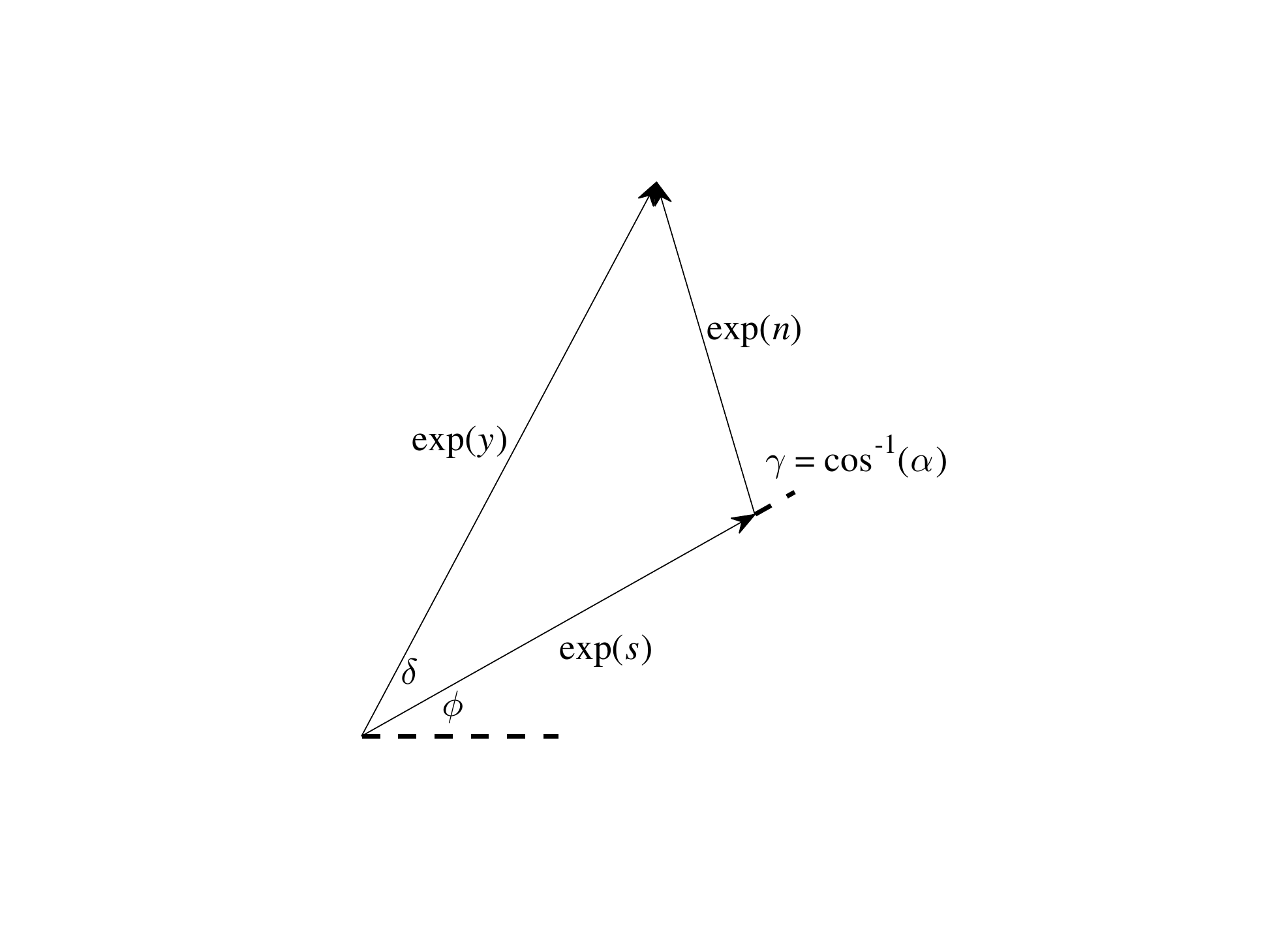}

\caption{Plot of the STFT phasors in the complex plane. }
\end{figure}

In this paper, we perform phase-sensitive speech enhancement using
a nonlinear modified KF in the log-spectral domain using speech phase
tracking. Speech and noise are taken to be additive in the complex
STFT domain, as in \cite{a225} \cite{a236}. Speech phase tracking
can be beneficial for enhancement because successive segments of speech
are correlated \cite{a19}. Modulation-domain Kalman filtering is
performed using speech phase inter-frame correlation modelling and
a phase-sensitive KF update that uses KF-based local priors for the
speech phase. In the KF update, the posterior distributions of the
speech phase and of the speech and noise log-spectra are computed
using a specific four-dimensional distribution parametrisation. 

The speech log-spectrum and phase and the noise log-spectrum are tracked
using one KF for each frequency bin. A KF is utilised to deal with
the temporal variability of speech: the temporal dynamics of the speech
log-spectrum and phase are modelled using the KF prediction step.
In the KF update, the estimated dynamics of the speech log-spectrum
and phase are combined with the observed noisy log-spectrum and phase
to obtain a minimum mean squared error (MMSE) estimate of the speech
log-spectrum and phase. The ordinary technique in order to incorporate
the speech phase is to track and estimate either the complex STFT
domain of speech or the real and imaginary parts of the STFT domain
of speech \cite{a13} \cite{a251}. 

The structure of this paper is as follows. Section II presents the
signal model, Sec. III describes the main phase-sensitive Kalman filtering
algorithm that performs speech phase estimation and Sec. IV introduces
the simplified versions of the main algorithm without speech phase
estimation. The algorithms are evaluated in Sec. V and conclusions
are drawn in Sec. VI. 

\section{Signal model and notation }

In the time domain, the noisy speech signal, $\tilde{y}(l)$, is given
by $\tilde{y}(l) = \tilde{s}(l) + \tilde{n}(l)$, where $\tilde{s}(l)$
and $\tilde{n}(l)$ are the clean speech and additive noise respectively
and $l$ is the discrete time index. Applying the STFT, we obtain
$Y_t(k)=S_t(k)+N_t(k)$ where $t$ indexes the time-frame and $k$
the frequency bin. Within the KF algorithm, frequency bins are processed
independently and, for clarity in the remainder of this paper, the
frequency index, $k$, is omitted and the time-frame index, $t$,
is included only in equations that involve multiple time-frames. 

In the complex STFT domain, we have: $Y=S+N \Leftrightarrow |Y|e^{j\theta}=|S|e^{j\phi}+|N|e^{j\psi}$
where the noisy speech phase is $\theta$, the speech phase is $\phi$
and the noise phase is $\psi$. The STFT spectral log-amplitudes of
the noisy speech, speech and noise are denoted by $y=\log|Y|$, $s=\log|S|$
and $n=\log|N|$. 

Taking the complex logarithm of the STFT coefficients gives $e^{y+j\theta}=e^{s+j\phi}+e^{n+j\psi}$.
We define the relative phases of the noisy speech and noise by $\delta=\theta-\phi$
and $\gamma=\psi-\phi$ respectively. Figure 1 depicts the STFT phasors
in the complex plane including the angles $\delta$ and $\gamma$.
Substituting for $\theta$ and $\psi$ gives $e^{y+j\delta}=e^{s}+e^{n+j\gamma}$.
We note that if $\delta,\gamma$ lie in the range $[-\pi, \pi)$ then,
since $e^{s}$ is real and positive, $\delta$ and $\gamma$ have
the same sign, $\text{sgn}(\delta)=\text{sgn}(\gamma)$; this will
be used in Sec. III.D. 

For convenience, we also define the phase factor, $\alpha=\cos \gamma$.
If $S$ and $N$ are independent, then $\gamma$ is uniform, $\gamma \sim U(-\pi,\pi)$,
and the phase factor distribution, $p(\alpha)$, is given by 

\begin{align}
 & p(\alpha)=\begin{cases}
\dfrac{1}{\pi\ \sqrt{1-\alpha^{2}}} & \ \ \text{if }-1<\alpha<1\\
0 & \ \ \text{otherwise.}
\end{cases}
\end{align}

Equation (1) is also utilised in \cite{a225}, \cite{a236} and \cite{a238}. 

In this paper, the semicolon ``;'' is used to denote vertical concatenation
of vectors: $(\textbf{x}_1; \ \textbf{x}_2) = (\textbf{x}_1^T \ \textbf{x}_2^T)^T$.
In addition, $\mathcal{Y}_t$ is used to denote all the noisy complex
STFT coefficients up to time $t$. Furthermore, the notation $s_{t|t}=\mathbb{E}\left\{ s_{t}\,|\,\mathcal{Y}_{t}\right\}$
is utilised to denote the posterior of a random variable, $s_t$. 

\subsection{Circular distributions and moments }

The distribution of a phase, $\phi$, is commonly modelled using either
the von Mises (vM) distribution or the Wrapped Normal (WN) distribution
which are both defined by two parameters. In both cases, the distibution
can be uniquely defined by its complex-valued first circular moment,
$E(\exp(j \phi))$ \cite{a215}. The vM distribution and its first
circular moment are given by 

\begin{align}
 & p_{\text{vM}}\left(\phi;\mu^{\text{\text{(\ensuremath{\phi})}}},\kappa^{\text{\text{(\ensuremath{\phi})}}}\right)=\dfrac{\exp\left(\kappa^{\text{\text{(\ensuremath{\phi})}}}\cos\left(\phi-\mu^{\text{\text{(\ensuremath{\phi})}}}\right)\right)}{2\pi\ I_{0}\left(\kappa^{\text{\text{(\ensuremath{\phi})}}}\right)}
\end{align}

and 

\begin{align}
 & \mathbb{E}\{\exp(j\phi)\}=\exp\left(j\mu^{\text{\text{(\ensuremath{\phi})}}}\right)\dfrac{I_{1}\left(\kappa^{\text{\text{(\ensuremath{\phi})}}}\right)}{I_{0}\left(\kappa^{\text{\text{(\ensuremath{\phi})}}}\right)}
\end{align}

respectively. In (2) and (3), $I_0(x)$ is the Bessel functions of
the order zero and, in (3), $I_1(x)$ is the Bessel function of the
order one \cite{a215}. In addition, in (2) and (3), $\mu^{\text{\text{($\phi$)}}}$
is the mean and $\kappa^{\text{\text{($\phi$)}}}$ is the concentration
of the vM distribution. We denote the vM distribution with mean $\mu_t^{\text{\text{($\phi$)}}}$
and concentration $\kappa_t^{\text{\text{($\phi$)}}}$ by vM($\mu_t^{\text{\text{($\phi$)}}}$,
$\kappa_t^{\text{\text{($\phi$)}}}$). For a given complex-valued
first trigonometric moment, $m_1=|m_1|e^{j \angle{m_1}}$, the vM
distribution with this first moment has the density $\smallskip$vM$( \angle{m_1}, \ \frac{I_0(|m_1|)}{I_1(|m_1|)} )$
\cite{a215}. 

The WN distribution and its first circular moment are 

\begin{align}
 & \ p_{\text{WN}}(\phi;\mu^{\text{\text{(\ensuremath{\phi})}}},\sigma^{2\text{{(\ensuremath{\phi})}}})\\
 & =\dfrac{1}{\sigma^{\text{{(\ensuremath{\phi})}}}\sqrt{2\pi}}\sum_{k=-\infty}^{\infty}\exp\left(\dfrac{-(\phi-\mu^{\text{\text{(\ensuremath{\phi})}}}+2\pi k)^{2}}{2\sigma^{2\text{{(\ensuremath{\phi})}}}}\right)\nonumber 
\end{align}

and 

\begin{align}
 & \mathbb{E}\{\exp(j\phi)\}=\exp\left(j\mu_{t}^{\text{\text{(\ensuremath{\phi})}}}-0.5\sigma_{t}^{2\text{{(\ensuremath{\phi})}}}\right)
\end{align}

respectively. In (4) and (5), $\mu_t^{\text{\text{($\phi$)}}}$ is
the mean and $\sigma_t^{2\text{{($\phi$)}}}$ is the variance of the
WN distribution. We denote the WN distribution with mean $\mu_t^{\text{\text{($\phi$)}}}$
and variance $\sigma_t^{2\text{{($\phi$)}}}$ by WN($\mu_t^{\text{{($\phi$)}}}$,
$\sigma_t^{2\text{\text{($\phi$)}}}$). For a given complex-valued
first trigonometric moment, $m_1=|m_1|e^{j \angle{m_1}}$, the WN
distribution with this first moment has the probability density WN($\angle{m_1}$,
$-2\log(|m_1|)$) \cite{a215}. 

\section{Description of the Enhancement Algorithm }

\subsection{Overview of the Algorithm }

The algorithm's flowchart is shown in Fig. 2. The core of the algorithm
is the KF enclosed in the figure by a dashed line. The KF tracks the
posterior distributions of the speech phase and of the log-spectra
of both the speech and the noise. Rather than tracking the speech
phase, $\phi$, directly, it is more convenient to track the unit-magnitude
complex quantity, $\exp(j \phi)$. 

The first step of the algorithm is to transform the noisy speech signal
into the time-frequency domain using the STFT. The algorithm considers
the spectral amplitude and phase of the noisy speech separately. The
noisy spectral amplitude is used in three ways: (a) it is converted
to the log-power domain and used as the KF observation together with
the noisy phase, $\theta$, (b) it is pre-cleaned and used for speech
AR modelling of order $p$ in the log-spectral domain, and (c) it
is used for noise AR modelling of order $q$ in the log-spectral domain.
The noisy phase is converted to the complex domain and then used for
speech phase complex AR modelling of order $1$. 

For the prediction step of the KF, autoregressive AR models are estimated
for $\exp(j \phi)$ and for the log-spectra of speech and noise. To
estimate the speech log-spectrum AR model, the noisy speech spectrum
is pre-cleaned using a conventional speech enhancer, converted to
the log-power domain and then divided into overlapping modulation
frames followed by AR analysis. To estimate the AR model for $\exp(j \phi)$,
the noisy phase is pre-cleaned using a speech phase enhancer, converted
to the complex domain and then divided into overlapping modulation
frames followed by AR analysis. 

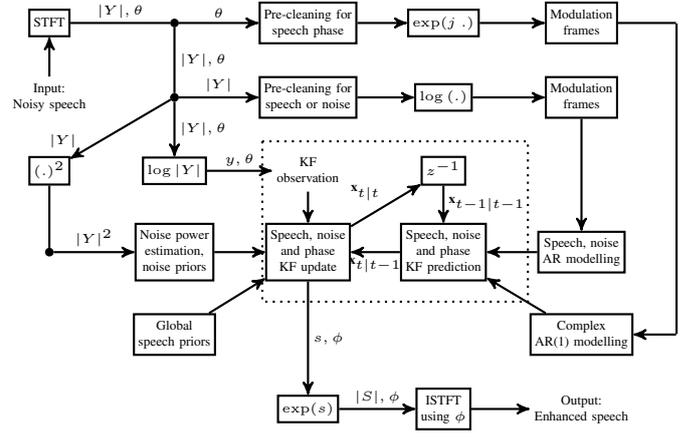
\begin{figure}[t]
\centering \begin{center}
  \begin{tikzpicture}[scale=0.22, auto, >=stealth']
    \tiny

    \matrix[ampersand replacement=\&, row sep=0.42cm, column sep=0.56cm] {
       \node[block] (g2) {$\begin{matrix} \text{STFT} \end{matrix}$}; \& \node[branch] (u1_81) {};
       \&  \node[block] (u1_82) {$\begin{matrix} \text{Pre-cleaning for} \\ \text{speech phase} \end{matrix}$};  \& \node[block] (u1_82_2) {$\begin{matrix} \exp (j \ .) \end{matrix}$};

       \&  \node[block] (u1_83) {$\begin{matrix} \text{Modulation} \\ \text{frames}  \end{matrix}$}; \& \node[branch2] (g2_01) {}; \\

       \node[] (g2hh3) {$\begin{matrix} \text{Input:} \\ \text{Noisy speech} \end{matrix}$}; \& \node[branch] (g2ii2) {}; 
       \& \node[block] (f1) {$\begin{matrix} \text{Pre-cleaning for} \\ \text{speech or noise} \end{matrix}$}; \& 

      \node[block] (f141) {$\begin{matrix} \log \left( . \right) \end{matrix}$}; \& \node[block] (f141t) {$\begin{matrix} \text{Modulation} \\ \text{frames}  \end{matrix}$}; \\

        \node[block] (myMY) {$\begin{matrix} \left( . \right)^2 \end{matrix}$}; \& \node[block] (F1_001) {$\begin{matrix} \log |Y| \end{matrix}$}; \& \node[] (v1) {$\begin{matrix} \text{KF}  \\ \text{observation} \end{matrix}$}; \& \node[block] (o441) {$\begin{matrix} z^{-1} \end{matrix}$}; \&    \\

       \node[branch] (F1_001_pp21) {}; \& \node[block] (g2998rr2) {$\begin{matrix} \text{Noise power} \\ \text{estimation,} \\ \text{noise priors} \end{matrix}$}; \& \node[block] (F155112) {$\begin{matrix} \text{Speech, noise} \\ \text{and phase} \\ \text{KF update} \end{matrix}$}; \&
      \node[block] (o1) {$\begin{matrix} \text{Speech, noise} \\ \text{and phase} \\ \text{KF prediction} \end{matrix}$}; \& \node[block] (f1411214) {$\begin{matrix} \text{Speech, noise} \\ \text{AR modelling} \end{matrix}$}; \\

      \& 
      \node[block] (g2998) {$\begin{matrix} \text{Global} \\ \text{speech priors} \end{matrix}$}; \&  \& \&  \node[block] (g2_03) {$\begin{matrix} \text{Complex} \\ \text{AR(1) modelling} \end{matrix}$};  \& \node[branch2] (g2_02) {}; \\

      \&  \& \node[block] (i2212211) {$\begin{matrix} \text{$\exp( s )$} \end{matrix}$}; \& \node[block] (i221) {$\begin{matrix} \text{ISTFT} \\ \text{using } \phi \end{matrix}$}; \& \node[] (i21) {$\begin{matrix} \text{Output:} \\ \text{Enhanced speech} \end{matrix}$}; \\
};

    \draw [connector] (f1) -- (f141);

\draw [connector] (g2ii2) -- node {\kern-1em \kern-1em \kern-1em \kern-2em \kern-2em \kern-2em $|Y|$} (myMY);

\draw [line] (myMY) -- (F1_001_pp21);

\draw [connector] (g2hh3) -- (g2);

\draw [connector] (g2998) -- (F155112); 
\draw [connector] (g2998rr2) -- (F155112); 

    \draw [line] (u1_81) -- node {$|Y|,\theta$} (g2ii2);
	\draw [connector] (g2ii2) -- node {$|Y|,\theta$} (F1_001);

\draw [connector] (F1_001_pp21) -- node {$|Y|^2$} (g2998rr2);

\draw [connector] (g2ii2) -- node {$|Y|$} (f1);

\draw [connector] (F1_001) -- node {$y,\theta$}  (v1);

\draw [line] (g2) -- node {$|Y|,\theta$} (u1_81);

\draw [connector] (u1_81) -- node {$\theta$} (u1_82);
\draw [connector] (u1_82) -- (u1_82_2);
\draw [connector] (u1_82_2) -- (u1_83);

\draw [line] (u1_83) -- (g2_01);
\draw [line] (g2_01) -- (g2_02);

\draw [connector] (g2_02) -- (g2_03);

\draw [connector] (g2_03) -- (o1);




\draw [connector] (i221) -- (i21);







\draw [connector] (f141) -- (f141t);


\draw [connector] (f141t) -- (f1411214);

    \draw [connector] (f1411214) -- (o1);
    \draw [connector] (o441) -- node {$\textbf{x}_{t-1|t-1}$} (o1);

	\draw [connector] (o1) -- node {$\textbf{x}_{t|t-1}$} (F155112); 

    \draw [connector] (v1) -- (F155112); 
    \draw [connector] (F155112) -- node {\kern2em \kern2em \kern2em $\textbf{x}_{t|t}$} (o441);

    \draw [connector] (i2212211) -- node {$|S|,\phi$} (i221);
	\draw [connector] (F155112) -- node {$s,\phi$} (i2212211);


\draw[thick,dotted] ($(v1.north west)+(-0.58,0.52)$)  rectangle ($(o1.south east)+(2.45,-1.22)$);


  \end{tikzpicture}
\end{center}

\caption{The flowchart of the algorithm is illustrated. The blocks in the dotted
rectangle constitute the KF that tracks the speech and noise log-spectra,
$(s,n)$, and the speech phase, $\phi$. The term $z^{-1}$ refers
to one-frame delay. }

\label{fig:EDCs-1-2-1-2-1-1-1-2-1-1-1-2-1-1-1-1-4} 
\end{figure}

The linear KF prediction step (described in Sec III.B) uses the results
of the AR analysis to predict the KF state in the the current frame,
$t$, from previous observations. 

The KF update (described in Sec. III.D) is nonlinear and combines
the output of the KF prediction with the observed noisy speech log-spectrum
and phase while incorporating additional prior distributions for the
speech and noise log-spectra. The additional prior distribution of
the noise is determined using an independent noise estimator. The
global speech prior distribution is obtained by multiplying an offline-trained
speech model by the estimated active speech level \cite{a102}. 

The final step of the algorithm is to extract the estimated phase
and log-power of the current frame from the KF state vector and to
combine these to generate the corresponding STFT coefficient. The
inverse STFT (ISTFT) is then used to reconstruct the enhanced speech
signal in the time domain. 

\subsection{The KF state and the KF prediction step }

In Fig. 2, within the dotted rectangle, the KF state vector, $\textbf{x}_t$,
has dimension $(p+q+1)$ and may be partitioned as 

\begin{align}
 & \textbf{x}_{t}=\left(\textbf{x}_{t}^{\text{(s)}};\ \textbf{x}_{t}^{\text{(n)}};\ x_{t}^{\text{(p)}}\right)
\end{align}

where 

\begin{align}
 & \textbf{x}_{t}^{\text{(s)}}=(s_{t}\ s_{t-1}\ \dots\ s_{t-p+1})^{T}\in\Re^{p}\\
 & \textbf{x}_{t}^{\text{(n)}}=(n_{t}\ n_{t-1}\ \dots\ n_{t-q+1})^{T}\in\Re^{q}\\
 & x_{t}^{\text{(p)}}=\exp(j\phi_{t})\in\mathbb{C}\text{.}
\end{align}

Based on equations (6)-(9), the speech KF state is $\textbf{x}_t^{\text{(s)}}$,
the noise KF state is $\textbf{x}_t^{\text{(n)}}$ and the phase KF
state is $x_t^{\text{(p)}}$. For the speech KF state, $\textbf{x}_t^{\text{(s)}}$,
the speech signal model is 

\begin{align}
 & \textbf{x}_{t|t-1}^{\text{(s)}}=\textbf{A}_{t}^{\text{(s)}}\left(\textbf{x}_{t-1|t-1}^{\text{(s)}}-\zeta_{t}^{\text{(s)}}\right)+\zeta_{t}^{\text{(s)}}+\textbf{w}_{t}^{\text{(s)}}
\end{align}

where 

\begin{align}
 & \textbf{A}_{t}^{\text{(s)}}=\left(\begin{array}{c}
-\textbf{a}_{t}^{\text{(s)}T}\\
\textbf{I}\ \ \textbf{0}
\end{array}\right)\in\Re^{p\times p}\\
 & \textbf{Q}_{t}^{\text{(s)}}=\left(\begin{array}{c}
\ \epsilon_{t}^{\text{(s)}}\ \ \textbf{0}^{T}\\
\textbf{0}\ \ \ \textbf{0}
\end{array}\right)\in\Re^{p\times p}\text{.}
\end{align}

Equation (10) is the linear KF prediction equation that is based on
AR modelling and on (11) and (12). 

In (10)-(12), the speech KF transition matrix is $\textbf{A}_t^{\text{(s)}}$,
the speech KF transition noise covariance matrix is $\textbf{Q}_t^{\text{(s)}}$
and the speech KF transition noise is $\textbf{w}_{t}^{\text{(s)}}$.
The speech KF transition noise, $\textbf{w}_{t}^{\text{(s)}}$, is
zero-mean Gaussian with covariance matrix $\textbf{Q}_t^{\text{(s)}}$.
In addition, the vector of speech AR coefficients is $\textbf{a}_t^{\text{(s)}} \in \Re^p$
and the variance of the transition noise is $\epsilon_{t}^{\text{(s)}}$.
The prior of the speech KF state is denoted by $\textbf{x}_{t|t-1}^{\text{(s)}}$
and the posterior of the speech KF state at time $(t-1)$ is denoted
by $\textbf{x}_{t-1|t-1}^{\text{(s)}}$. 

In (10), $\textbf{A}_t^{\text{(s)}}$, $\zeta_t^{\text{(s)}}$ and
$\textbf{Q}_t^{\text{(s)}}$ are estimated from AR($p$) modelling
described in Sec III.G. The AR mean, $\zeta_t^{\text{(s)}}$, is the
mean speech log-amplitude estimated as a speech AR parameter. 

In (10), the way in which the speech KF state, $\textbf{x}_t^{\text{(s)}}$,
changes in the speech KF prediction step is defined. The speech KF
state consists of a speech KF state mean, $\pmb{\mu}_t^{\text{(s)}} \in \Re^p$,
and a speech KF state covariance matrix, $\textbf{P}_t^{\text{(s)}} \in \Re^{p \times p}$.
Both $\pmb{\mu}_t^{\text{(s)}}$ and $\textbf{P}_t^{\text{(s)}}$
are updated in the linear speech KF prediction using 

\begin{align}
 & \pmb{\mu}_{t|t-1}^{\text{(s)}}=\textbf{A}_{t}^{\text{(s)}}\left(\pmb{\mu}_{t-1|t-1}^{\text{(s)}}-\zeta_{t}^{\text{(s)}}\right)+\zeta_{t}^{\text{(s)}}\\
 & \textbf{P}_{t|t-1}^{\text{(s)}}=\textbf{A}_{t}^{\text{(s)}}\textbf{P}_{t-1|t-1}^{\text{(s)}}\textbf{A}_{t}^{\text{(s)}T}+\textbf{Q}_{t}^{\text{(s)}}\text{.}
\end{align}

For the noise KF state, $\textbf{x}_t^{\text{(n)}}$, the inter-frame
correlation between adjacent noise frames is modelled in the log-spectral
domain using a noise KF prediction step that is based on noise AR($q$)
modelling described in Sec III.G. As in (13) and (14), the noise KF
prediction step equations are 

\begin{align}
 & \pmb{\mu}_{t|t-1}^{\text{(n)}}=\textbf{A}_{t}^{\text{(n)}}\left(\pmb{\mu}_{t-1|t-1}^{\text{(n)}}-\zeta_{t}^{\text{(n)}}\right)+\zeta_{t}^{\text{(n)}}\\
 & \textbf{P}_{t|t-1}^{\text{(n)}}=\textbf{A}_{t}^{\text{(n)}}\textbf{P}_{t-1|t-1}^{\text{(n)}}\textbf{A}_{t}^{\text{(n)}T}+\textbf{Q}_{t}^{\text{(n)}}
\end{align}

where the quantities with the superscript (n) refer to the noise KF
state and are analogous to the speech quantities. 

The algorithm uses a joint speech and noise KF state with a full covariance
matrix, hence tracking the correlation between speech and noise in
the log-spectral domain \cite{a225}. The joint speech and noise KF
state is $\textbf{x}_{t}^{\text{(j)}}=(\textbf{x}_{t}^{\text{(s)}};\ \textbf{x}_{t}^{\text{(n)}})\in\Re^{p+q}$.
The superscript (j) denotes the joint KF state. For the joint KF state,
$\textbf{x}_{t}^{\text{(j)}}$, the joint speech and noise signal
model is 

\begin{align}
 & \textbf{x}_{t|t-1}^{\text{(j)}}=\textbf{A}_{t}^{\text{(j)}}\left(\textbf{x}_{t-1|t-1}^{\text{(j)}}-\pmb{\zeta}_{t}^{\text{(j)}}\right)+\textbf{w}_{t}^{\text{(j)}}+\pmb{\zeta}_{t}^{\text{(j)}}
\end{align}

where 

\begin{align}
 & \textbf{A}_{t}^{\text{(j)}}=\left(\begin{array}{c}
\textbf{A}_{t}^{\text{(s)}}\ \ \ \textbf{0}\\
\textbf{0}\ \ \ \textbf{A}_{t}^{\text{(n)}}
\end{array}\right)\in\Re^{(p+q)\times(p+q)}\\
 & \pmb{\zeta}_{t}^{\text{(j)}}=(\zeta_{t}^{\text{(s)}};\ \dots;\ \zeta_{t}^{\text{(s)}};\ \zeta_{t}^{\text{(n)}};\ \dots;\ \zeta_{t}^{\text{(n)}})\in\Re^{(p+q)}\\
 & \textbf{Q}_{t}^{\text{(j)}}=\left(\begin{array}{c}
\textbf{Q}_{t}^{\text{(s)}}\ \ \ \textbf{0}\\
\textbf{0}\ \ \ \textbf{Q}_{t}^{\text{(n)}}
\end{array}\right)\in\Re^{(p+q)\times(p+q)}\text{.}
\end{align}

The KF prediction step update equations for the mean, $\pmb{\mu}_{t}^{\text{(j)}}$,
and covariance matrix, $\textbf{P}_{t}^{\text{(j)}}$, are therefore 

\begin{align}
 & \pmb{\mu}_{t|t-1}^{\text{(j)}}=\textbf{A}_{t}^{\text{(j)}}\left(\pmb{\mu}_{t-1|t-1}^{\text{(j)}}-\pmb{\zeta}_{t}^{\text{(j)}}\right)+\pmb{\zeta}_{t}^{\text{(j)}}\\
 & \textbf{P}_{t|t-1}^{\text{(j)}}=\textbf{A}_{t}^{\text{(j)}}\textbf{P}_{t-1|t-1}^{\text{(j)}}\textbf{A}_{t}^{\text{(j)}T}+\textbf{Q}_{t}^{\text{(j)}}\\
 & \pmb{\mu}_{t|t}^{\text{(j)}}=\left(\pmb{\mu}_{t|t}^{\text{(s)}};\ \pmb{\mu}_{t|t}^{\text{(n)}}\right)\in\Re^{(p+q)}\\
 & \textbf{P}_{t|t}^{\text{(j)}}=\left(\begin{array}{c}
\textbf{P}_{t|t}^{\text{(s)}}\ \ \ \textbf{Z}^{T}\\
\textbf{Z}\ \ \ \textbf{P}_{t|t}^{\text{(n)}}
\end{array}\right)\in\Re^{(p+q)\times(p+q)}
\end{align}

where $\textbf{P}_{t}^{\text{(j)}}$ can be decomposed as shown in
the last line. 

For the speech phase KF state, $x_t^{\text{(p)}}$, $\exp(j \phi_t)$
is utilised. Using $\exp(j \phi_t)$ as the speech phase KF state
leads to $E \{ \exp(j \phi) \exp(-j \phi) \} = 1$ and, in turn, this
leads to (25). The speech phase KF state variance, $P_t^{\text{(p)}} \in \Re$,
is related to the speech phase KF state mean, $\mu_{t}^{\text{(p)}} \in \mathbb{C}$,
with 

\begin{align}
 & P_{t}^{\text{(p)}}=1-|\mu_{t}^{\text{(p)}}|^{2}\text{.}
\end{align}

Due to (25), the algorithm tracks the phase KF state mean, $\mu_{t}^{\text{(p)}}$,
and does not utilise $P_t^{\text{(p)}}$. In (6), the KF state, $\textbf{x}_t \in \mathbb{C}^{p+q+1}$,
is defined and, now, the KF state mean is 

\begin{align}
 & \pmb{\mu}_{t|t-1}=(\pmb{\mu}_{t|t-1}^{\text{(j)}};\ \mu_{t|t-1}^{\text{(p)}})\text{.}
\end{align}

It should be noted that $\pmb{\mu}_{t}$ and $\textbf{P}_{t}^{\text{(j)}}$
do not correspond to each other. According to (26), $\pmb{\mu}_{t}$
includes both $\pmb{\mu}_{t|t-1}^{\text{(j)}}$ and $\mu_{t}^{\text{(p)}}$. 

\begin{figure}[t]
\centering \begin{center}
  \begin{tikzpicture}[scale=0.42, auto, >=stealth']
    \tiny

    \matrix[ampersand replacement=\&, row sep=0.04cm, column sep=0.63cm] {
\&  \node[] (f1r2) {$\begin{matrix} \text{From complex AR($1$)} \\ \text{modelling: } {A}_t^{\text{(p)}}, {\epsilon}_t^{\text{(p)}} \end{matrix}$}; \&
       \&  

      \node[] (f1r) {$\begin{matrix} y,\theta \text{ and} \\ (s,n) \text{ priors} \end{matrix}$}; \& \&  \\

\node[] (i14141r122) {$\begin{matrix} \text{${x}_{t-1|t-1}^{\text{(p)}}$} \smallskip \\ \text{${\mu}_{t-1|t-1}^{\text{(p)}}$} \in \smallskip \mathbb{C} \end{matrix}$}; \& \node[block] (f1lllcc22) {$\begin{matrix} \text{Phase KF} \\ \text{prediction} \end{matrix}$}; \& 
       \node[] (u1t22) {$\begin{matrix} {x}_{t|t-1}^{\text{(p)}} \smallskip \\ \text{${\mu}_{t|t-1}^{\text{(p)}}$} \end{matrix}$}; \& 

      \node[block] (f1tuu) {$\begin{matrix} \text{Phase KF} \\ \text{update} \end{matrix}$}; \& \node[] (f141tuu) {$\begin{matrix} \text{${x}_{t|t}^{\text{(p)}}$}  \smallskip \\ {\mu}_{t|t}^{\text{(p)}} \end{matrix}$}; \& \\
    };

\draw [connector] (f1r2) -- (f1lllcc22);
\draw [connector] (f1r) -- (f1tuu);

\draw [connector] (i14141r122) -- (f1lllcc22);
\draw [connector] (f1lllcc22) -- (u1t22);
\draw [connector] (u1t22) -- (f1tuu);
\draw [connector] (f1tuu) -- (f141tuu);



 \end{tikzpicture}
\end{center}

\caption{The speech phase KF is shown. The inputs and outputs in this diagram
match the inputs and outputs of the dotted rectangle in Fig. 2 only
for the speech phase KF. The speech phase KF state is $\text{$x_{t}^{\text{(p)}}$}=\exp(j \phi_t)$. }

\label{fig:EDCs-1-2-1-2-1-1-1-2-1-1-1-2-1-1-1-1-1-1-1-2-1-1-2} 
\end{figure}
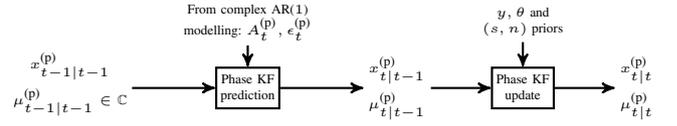

Figure 3 shows how the speech phase KF state, $\text{${x}_{t}^{\text{(p)}}$}$,
evolves over time. The three main points of Fig. 3 are: (a) only $\mu_{t}^{\text{(p)}}$
is used to track $\text{${x}_{t}^{\text{(p)}}$}$, (b) $\text{${x}_{t|t-1}^{\text{(p)}}$}$
is computed from $\text{${x}_{t-1|t-1}^{\text{(p)}}$}$ and from the
two parameters of complex AR($1$) modelling, and (c) the speech phase
posterior, $\text{${x}_{t|t}^{\text{(p)}}$}$, given the noisy $y$
and $\theta$ is calculated using the correlated log-spectrum local
priors of $(s,n)$. Regarding (b), the two parameters of the complex
AR($1$) modelling are the AR coefficient, $A_{t}^{\text{(p)}}\in\mathbb{C}$,
and the speech phase KF transition noise $\smallskip$variance, $\ \epsilon_{t}^{\text{(p)}} \in \Re$. 

Equations (27)-(30) show how $\mu_{t}^{\text{(p)}}$ changes in the
speech phase KF prediction step. To perform the KF prediction step,
$\check{\mu}_{t|t-1}^{\text{(p)}}$ is first obtained from the AR
model. Moment matching is then used to map this to a WN distribution
with parameters $\check{\mu}_{t|t-1}^{(\phi)}$ and $\check{\sigma}_{t|t-1}^{2(\phi)}$.
Next, the transition noise variance, $\epsilon_{t}^{\text{(p)}}$,
is added onto $\check{\sigma}_{t|t-1}^{2(\phi)}$ and is then converted
back to the first circular moment, $\mu_{t|t-1}^{\text{(p)}}$. These
steps are summarised as: 

\begin{align}
 & \check{\mu}_{t|t-1}^{\text{(p)}}=A_{t}^{\text{(p)}}\mu_{t-1|t-1}^{\text{(p)}}\\
 & \check{\mu}_{t|t-1}^{\text{(\ensuremath{\phi})}}=\text{atan2}\left(\Im\{\check{\mu}_{t|t-1}^{\text{(p)}}\},\Re\{\check{\mu}_{t|t-1}^{\text{(p)}}\}\right)\\
 & \check{\sigma}_{t|t-1}^{2\text{\text{(\ensuremath{\phi})}}}=-2\log(|\check{\mu}_{t|t-1}^{\text{(p)}}|)\\
 & \mu_{t|t-1}^{\text{(p)}}=\exp\left(j\check{\mu}_{t|t-1}^{\text{(\ensuremath{\phi})}}-0.5\left(\check{\sigma}_{t|t-1}^{2\text{\text{(\ensuremath{\phi})}}}+\epsilon_{t}^{\text{(p)}}\right)\right)\text{.}
\end{align}

Considering the uncertainty in the complex AR prediction, the transition
noise variance, $\epsilon_{t}^{\text{(p)}}$, reduces the absolute
value of $\mu_{t}^{\text{(p)}}$. Equations (27)-(30) are based on
the WN distribution and on trigonometric moment matching in Sec. II.A. 

Full covariance matrices are not used to model the correlation between
the speech phase, $\phi$, and the speech and noise log-spectra, $(s,n)$.
The algorithm assumes $p(s,n,\phi)= p(s,n) \ p(\phi)$ and thus that
$\phi$ is independent of $(s,n)$. 

\subsection{The nonlinear KF update step }

The algorithm performs the update $\textbf{x}_{t|t} = (\textbf{x}_{t|t}^{\text{(j)}};\ x_{t|t}^{\text{(p)}})$.
The Kalman filtering algorithm first estimates $x_{t|t}^{\text{(p)}}$
using 

\begin{align}
 & \mu_{t|t}^{\text{(p)}}=\mathbb{E}\left\{ x_{t}^{\text{(p)}}\,|\,\mathcal{Y}_{t}\right\} 
\end{align}

and it then computes $\textbf{x}_{t|t}^{\text{(j)}}$. 

To perform the KF update, the KF state vector, $\textbf{x}_{t|t-1}$,
is first transformed so that the elements corresponding to the current
time, $(s_t; \ n_t)$, are uncorrelated with the other elements \cite{a123}.
The current-frame elements, $(s_t; \ n_t)$, are decorrelated in the
KF state because the proposed KF update is nonlinear. Then, the mean
and the covariance matrix of these elements are updated as described
in Sec III.D. Finally, the transformation is reversed in order to
restore the original KF state vector. 

In the KF prediction presented in Sec. III.B, the priors, $\pmb{\mu}_{t|t-1}^{\text{(j)}}$
and $\textbf{P}_{t|t-1}^{\text{(j)}}$, are calculated in (23) and
(24). In the KF update in Sec. III.D, the posteriors, $\breve{\pmb{\mu}}_{t|t}^{\text{(j)}}$
and $\breve{\textbf{P}}_{t|t}^{\text{(j)}}$, are found. This section
shows how the decorrelation and recorrelation steps are used to calculate
$\pmb{\mu}_{t|t}^{\text{(j)}}$ and $\textbf{P}_{t|t}^{\text{(j)}}$. 

Before performing the KF update, the KF state is decorrelated by multiplying
the speech and noise state vector, $\textbf{x}^{\text{(j)}}$, by
the matrix $\textbf{D}$. The transformation matrix, $\textbf{D}$,
is chosen to decorrelate the current frame elements, $s_t$ and $n_t$,
from the rest of the state vector. Transformed quantities are identified
with a breve diacritic. The multiple-element decorrelation operation
preserves the inter-frame correlation that is created from the KF
prediction step so that the algorithm performs the KF update with
the variables in the current time-frame. 

The covariance matrix, $\textbf{P}_{t|t-1}$, is first decomposed
as 

\begin{align}
 & \textbf{P}_{t|t-1}^{\text{(j)}}=\left(\begin{array}{c}
\textbf{P}_{A}\ \ \ \ \ \ \ \textbf{P}_{B}^{T}\\
\textbf{P}_{B}\ \ \ \ \ \ \ \textbf{P}_{C}
\end{array}\right)\text{.}
\end{align}

Then, $\textbf{D}=\textbf{C}\textbf{B}$ is utilised where 

\begin{align}
 & \textbf{C}=\left(\begin{array}{c}
\textbf{I}\ \ \ \ \ \ \ \ \ \ \ \ \ \textbf{0}\\
-\textbf{P}_{B}\textbf{P}_{A}^{-T}\ \ \textbf{I}
\end{array}\right)
\end{align}

and where the permutation matrix $\textbf{B}\in\Re^{(p+q)\times(p+q)}$
is 

\begin{align}
 & \textbf{B}=[e_{1}\ e_{p+1}\ e_{2}\ e_{3}\ \dots\ e_{p}\ e_{p+2}\ e_{p+3}\ \dots\ e_{p+q}]^{T}
\end{align}

where $e_i$ is the $i$-th column of the identity matrix. Next, $\textbf{D}$
is used along with the linear transformation 

\begin{align}
 & \breve{\pmb{\mu}}_{t|t-1}^{\text{(j)}}=\textbf{D}\ \pmb{\mu}_{t|t-1}^{\text{(j)}}\\
 & \breve{\textbf{P}}_{t|t-1}^{\text{(j)}}=\textbf{D}\ \textbf{P}_{t|t-1}^{\text{(j)}}\ \textbf{D}^{T}\triangleq\left(\begin{array}{c}
\textbf{P}_{A}\ \ \ \ \ \ \ \ \ \ \ \ \ \ \ \ \ \ \ \textbf{0}\\
\textbf{0}\ \ \ \ \textbf{P}_{C}-\textbf{P}_{B}\textbf{P}_{A}^{-1}\textbf{P}_{B}^{T}
\end{array}\right)
\end{align}

where the off-diagonal block matrices are zero matrices. The decorrelation
step is the linear transformation in (35),$\,$(36). 

For convenience, from (36), we also define 

\begin{align}
 & \breve{\textbf{P}}_{\text{C}}=\textbf{P}_{C}-\textbf{P}_{B}\textbf{P}_{A}^{-1}\textbf{P}_{B}^{T}\text{.}
\end{align}

In (32) and (36), we observe that $\textbf{P}_A$ is preserved after
the multiple-element decorrelation operation. This is important since
$\textbf{P}_A$ is updated in the KF update in Sec. III. D. After
the KF update, the inverse transformation is applied with 

\begin{align}
 & \pmb{\mu}_{t|t}^{\text{(j)}}=\textbf{D}^{-1}\ \breve{\pmb{\mu}}_{t|t}^{\text{(j)}}\\
 & \textbf{P}_{t|t}^{\text{(j)}}=\textbf{D}^{-1}\ \breve{\textbf{P}}_{t|t}^{\text{(j)}}\ (\textbf{D}^{T})^{-1}\text{.}
\end{align}

Equations (38) and (39) constitute the recorrelation step. 

In (38) and (39), the posteriors $\breve{\pmb{\mu}}_{t|t}^{\text{(j)}}$
and $\breve{\textbf{P}}_{t|t}^{\text{(j)}}$ are computed using the
equations in (40) and (41). Now, $\breve{\pmb{\mu}}_{t|t}^{\text{(j)}}$
is defined by: 

\begin{align}
 & \breve{\pmb{\mu}}_{t|t}^{\text{(j)}}=\left(s_{t|t};\ n_{t|t};\ \breve{\pmb{\mu}}_{\text{C},t|t-1}^{\text{(j)}}\right),\ \breve{\pmb{\mu}}_{\text{C},t|t-1}^{\text{(j)}}\in\Re^{(p+q-2)}\\
 & \ \ \ \text{where }\ \breve{\pmb{\mu}}_{t|t-1}^{\text{(j)}}=\left(s_{t|t-1};\ n_{t|t-1};\ \breve{\pmb{\mu}}_{\text{C},t|t-1}^{\text{(j)}}\right)\nonumber 
\end{align}

where the $s_{t|t}$ notation is introduced in Sec. II. Furthermore,
the posterior covariance matrix, $\breve{\textbf{P}}_{t|t}^{\text{(j)}}$,
is defined by: 

\begin{align}
 & \breve{\textbf{P}}_{t|t}^{\text{(j)}}=\left(\begin{array}{c}
\breve{\textbf{P}}_{\text{A},t|t}\ \ \ \ \ \ \ \ \ \textbf{0}\\
\textbf{0}\ \ \ \ \ \ \ \breve{\textbf{P}}_{\text{C},t|t-1}
\end{array}\right)\\
 & \ \ \ \text{where }\ \breve{\textbf{P}}_{t|t-1}^{\text{(j)}}=\left(\begin{array}{c}
\breve{\textbf{P}}_{\text{A},t|t-1}\ \ \ \ \ \ \ \ \textbf{0}\\
\textbf{0}\ \ \ \ \ \ \ \ \breve{\textbf{P}}_{\text{C},t|t-1}
\end{array}\right)\nonumber \\
 & \ \ \ \breve{\textbf{P}}_{\text{A},t|t}\in\Re^{2\times2},\ \ \ \breve{\textbf{P}}_{\text{C},t|t-1}\in\Re^{(p+q-2)\times(p+q-2)}\nonumber \\
 & \ \ \ \breve{\textbf{P}}_{\text{A},t|t}=\left(\begin{array}{c}
\mathbb{E}\left\{ s_{t}^{2}\,|\,\mathcal{Y}_{t}\right\} \ \ \ \ \mathbb{E}\left\{ s_{t}n_{t}\,|\,\mathcal{Y}_{t}\right\} \\
\mathbb{E}\left\{ s_{t}n_{t}\,|\,\mathcal{Y}_{t}\right\} \ \ \mathbb{E}\left\{ n_{t}^{2}\,|\,\mathcal{Y}_{t}\right\} 
\end{array}\right)\nonumber \\
 & \ \ \ \ \ \ \ \ \ \ \ \ -\left(s_{t|t};\ n_{t|t}\right)\left(s_{t|t};\ n_{t|t}\right)^{T}\text{.}\nonumber 
\end{align}

In Sec III.D, $\mathbb{E}\{(s_t; n_t; x_{t}^{\text{(p)}}) \ | \ \mathcal{Y}_{t}\}$
is calculated along with the covariance matrix of the first two elements
of this vector. 

\subsection{The phase-sensitive modified KF update }

This section describes the evaluation of the quantities needed to
perform the KF update step described in Sec II.C, namely the posterior
distribution parameters $(s_{t|t}; n_{t|t}; \mu^{\text{(p)}}_{t|t})$
in (40) and (31) together with the speech and noise covariance matrix
$\breve{\textbf{P}}_{t|t}^{\text{(j)}}$ in (41). The algorithm tracks
the complex-valued speech phase posterior mean, $\mu_{t|t}^{\text{(p)}}$,
without its real-valued variance due to $\pmb{\mu}_{t|t}=(\pmb{\mu}_{t}^{\text{(j)}}; \ \mu_{t}^{\text{(p)}})$
in (26). 

In this section, the time subscript, $t$, is omitted from all the
random variables for clarity, as described in Sec. II. 

The prior distribution, $p(s,n,\phi,\psi)=p(s,n)\ p(\phi)\ p(\psi)$,
may be obtained from the KF prediction step outputs, $\pmb{\mu}_{t|t-1}$
and $\textbf{P}_{t|t-1}^{\text{(j)}}$, together with the assumption
that $\psi$ is uniformly distributed, $\psi \sim U(-\pi, \pi)$.
To impose the observation constraint, we use the relationship $e^{y+j\delta}=e^{s}+e^{n+j\gamma}$
illustrated by the STFT phasor diagram in Fig. 1. We assume that $p(s,n,\phi,\psi)$
can be decomposed as the product of three independent distributions
over the domain $-\infty<s,\,n<\infty$ and $-\pi<\phi,\,\psi\le\pi$.
To calculate the conditional distribution for a given observation,
$y_{t}$ and $\theta_{t}$, we make a change of variables and we parametrise
the distribution using $(u,\,y,\,\gamma,\,\theta)$. This invertible
and one-to-one transformation is given by: 

\begin{eqnarray}
 &  & \hspace*{-0.6cm}u=n-s\\
 &  & \hspace*{-0.6cm}y=0.5\left(s+n+\log\left(2\cosh(u)+2\cos\gamma\right)\right)\\
 &  & \hspace*{-0.6cm}\theta=\phi+\delta\\
 &  & \hspace*{-0.6cm}\ \,\,\,=\phi+\mathrm{sgn}\left(\gamma\right)\cos^{-1}\left(\frac{e^{2y}+e^{2s}-e^{2n}}{2e^{s+y}}\right)\nonumber \\
 &  & \hspace*{-0.6cm}\ \,\,\,=\phi+\mathrm{sgn}\left(\gamma\right)\cos^{-1}\left(\cosh\left(y-s\right)-0.5e^{2n-s-y}\right)\text{.}\nonumber 
\end{eqnarray}

Equations (42)-(44) are over the domain $-\infty<u,\,y<\infty$ and
$-\pi<\gamma,\,\theta\le\pi$, assuming $0\le\cos^{-1}\left(.\right)\le\pi$. 

Equation (43) is the nonlinear log-spectral distortion equation, which
is also in \cite{a129}. Equation (43) can be derived by multiplying
$e^{y+j\delta}=e^{s}+e^{n+j\gamma}$ by its complex conjugate and
taking logs. Multiplying $e^{y+j\delta}=e^{s}+e^{n+j\gamma}$ by its
complex conjugate gives $e^{2y}=e^{s+n}\left(e^{s-n}+e^{n-s}+2\cos\gamma\right)$. 

In addition, equation (44) can be derived by writing $e^{y+j\theta}-e^{s+j\phi}=e^{n+j\psi}$
and multiplying this equation by its complex conjugate, which gives
$e^{2y}+e^{2s}-2e^{s+y}\cos\delta=e^{2n}$. 

The inverse transformation of (42)-(44) is given by: 

\begin{eqnarray}
 &  & \hspace*{-0.6cm}v=s+n=2y-\log\left(2\cosh(u)+2\cos\gamma\right)\\
 &  & \hspace*{-0.6cm}s=0.5\left(v-u\right),\ \ \ \ \ \ n=0.5\left(v+u\right)\\
 &  & \hspace*{-0.6cm}\delta=\mathrm{sgn}\left(\gamma\right)\\
 &  & \hspace*{-0.6cm}\ \ \ \ \ \times\cos^{-1}\left(\cosh\left(y-s\right)-0.5e^{2n-s-y}\right)\nonumber \\
 &  & \hspace*{-0.6cm}\phi=\theta-\delta,\ \ \ \ \ \ \psi=\gamma+\phi\text{.}
\end{eqnarray}

We use $(s,n,\phi,\psi) \Rightarrow (u,y,\gamma,\theta)$. The Jacobian
matrix, $\frac{\partial(u,\,y,\,\gamma,\,\theta)}{\partial(s,\,n,\,\phi,\,\psi)}$,
is computed and the Jacobian is $\Delta=1$. 

Now, we form the conditional distribution $p(u,\gamma | y, \theta)$
to apply the observation constraint. To calculate the posterior mean
and covariance matrices, $\breve{\pmb{\mu}}_{t|t}^{\text{(j)}}$ and
$\breve{\textbf{P}}_{t|t}^{\text{(j)}}$, that are required in (38)
and (39), we compute expectations of the form: 

\begin{eqnarray}
 &  & \hspace*{-0.6cm}\mathbb{E}\left\{ f\left(.\right)\,|\,y_{t},\theta_{t}\right\} \ =\ \int_{u,\gamma}\ f(.)\ p(u,\gamma|y_{t},\theta_{t})\ du\ d\gamma\nonumber \\
 &  & \hspace*{-0.6cm}\propto\ \int_{u,\gamma}\ f(.)\ p(u,\gamma,y_{t},\theta_{t})\ du\ d\gamma\\
 &  & \hspace*{-0.6cm}=\ \int_{u,\gamma}\ f(.)\ |\Delta|^{-1}\ p(s,n,\phi,\psi)\ du\ d\gamma\nonumber 
\end{eqnarray}

\begin{eqnarray*}
 &  & \hspace*{-0.6cm}\propto\ \int_{u,\gamma}\ f(.)\ p(s,n,\phi,\psi)\ du\ d\gamma\\
 &  & \hspace*{-0.6cm}=\ \int_{u,\gamma}\ f(.)\ p(s,n)\ p(\phi)\ p(\psi)\ du\ d\gamma
\end{eqnarray*}

where $(s,n,\phi,\psi)$ are obtained from $(u, \gamma, y_t, \theta_t)$
using the equations in (45)-(48). In (49), the speech phase, $\phi$,
is assumed to be independent of the speech and noise log-powers. 

Based on Sec. II, $\psi \sim U(-\pi,\pi)$ and thus (49) becomes: 

\begin{eqnarray}
 &  & \hspace*{-0.6cm}\hspace*{-0.6cm}\mathbb{E}\left\{ f\left(.\right)\,|\,y_{t},\theta_{t}\right\} \propto\int_{u,\gamma}\ f(.)\ p(s,n)\ p(\phi)\ du\ d\gamma\text{.}
\end{eqnarray}

In (49)-(50), we compute conditional expectations and we use $f(.)$
to denote any function $f(u, y, \gamma, \theta)$. We can now calculate
a conditional expectation and we can then split the integral into
two regions according to the sign of $\gamma$: 

\begin{eqnarray}
 &  & \hspace*{-0.6cm}\hspace*{-0.6cm}\mathbb{E}\left\{ f(.)|y_{t},\theta_{t}\right\} \propto\int_{u,\gamma}\ f(.)\ p(s,n)\ p(\phi)\ du\ d\gamma\\
 &  & \hspace*{-0.6cm}\hspace*{-0.6cm}=\int_{-\pi}^{0}\int_{u}\ f\left(.\right)\ p(s,n)\ p(\phi|\gamma\leq0)\ du\ d\gamma\nonumber \\
 &  & \hspace*{-0.6cm}\hspace*{-0.6cm}+\int_{0}^{\pi}\int_{u}\ f\left(.\right)\ p(s,n)\ p(\phi|\gamma>0)\ du\ d\gamma\text{.}\nonumber 
\end{eqnarray}

We now make the one-to-one transformation $\alpha=\cos\gamma$ separately
for each of the two integrals in (51): 

\begin{eqnarray}
 &  & \hspace*{-0.6cm}\mathbb{E}\left\{ f\left(.\right)\,|\,y_{t},\theta_{t}\right\} \\
 &  & \hspace*{-0.6cm}=\int_{\alpha}\int_{u}\ f\left(.\right)\ p(s,n)\ p(\phi|\gamma\leq0)\ |\Delta_{\alpha}|^{-1}\ du\ d\alpha\nonumber \\
 &  & \hspace*{-0.6cm}+\int_{\alpha}\int_{u}\ f\left(.\right)\ p(s,n)\ p(\phi|\gamma>0)\ |\Delta_{\alpha}|^{-1}\ du\ d\alpha\nonumber 
\end{eqnarray}

where both integrals are over the domain $-\infty<u<\infty$ and $-1<\alpha<1$.
In (52), \smallskip$\Delta_{\alpha} = \frac{d \alpha}{d \gamma} = -\sqrt{1-\alpha^2}$. 

Using (52), the first and second moments of $(s,n, x_t^{\text{(p)}})$
are computed. Due to $p(s,n,\phi)= p(s,n)\times p(\phi)$, the correlations
between $s$ and $\phi$ and between $n$ and $\phi$ are always zero
in the KF prior distribution but are not zero in the KF \smallskip
posterior. 

Using $0 \leq a+b \leq 2$, where $a$ and $b$ are non-negative integers,
we now compute the posterior for $(s,n)$. We utilise $p(\alpha)$
as defined in (1). The posterior for $(s,n)$ is based on: 

\begin{eqnarray}
 &  & \hspace*{-1cm}\mathbb{E}\left\{ s^{a}n^{b}\,|\,y_{t},\,\theta_{t}\right\} \propto\int_{\alpha}\dfrac{1}{\sqrt{1-\alpha^{2}}}\nonumber \\
 &  & \hspace*{-1cm}\times\int_{u}s^{a}n^{b}\times p(u,v)p(\phi|\gamma\leq0)\ du\ d\alpha\\
 &  & \hspace*{-1cm}+\int_{\alpha}\dfrac{1}{\sqrt{1-\alpha^{2}}}\times\int_{u}s^{a}n^{b}\times p(u,v)p(\phi|\gamma>0)\ du\ d\alpha\nonumber 
\end{eqnarray}

where $v$ is defined in (45) based on $y$, $u$ and $\alpha=\cos \gamma$.
In (53), $p(\phi)$ is utilised along with the mapping from $E\{x_t^{\text{(p)}}\}$
to $p(\phi)$ using the vM distribution presented in Sec. II.A. 

Moreover, using equation (52), in order to compute the posterior for
$x_t^{\text{(p)}}$, we now compute: 

\begin{eqnarray}
 &  & \ \ \ \hspace*{-1cm}\mathbb{E}\left\{ x_{t}^{\text{(p)}}\,|\,y_{t},\,\theta_{t}\right\} \propto\int_{\alpha}\dfrac{1}{\sqrt{1-\alpha^{2}}}\nonumber \\
 &  & \ \ \ \hspace*{-1cm}\int_{u}\exp(j\phi|\gamma\leq0)\ p(u,v)p(\phi|\gamma\leq0)\ du\ d\alpha+\\
 &  & \ \ \ \hspace*{-1cm}\int_{\alpha}\dfrac{1}{\sqrt{1-\alpha^{2}}}\ \int_{u}\exp(j\phi|\gamma>0)\ p(u,v)p(\phi|\gamma>0)\ du\ d\alpha\text{.}\nonumber 
\end{eqnarray}

Equations (53) and (54) provide the link between Sec. III.C and Sec.
III.D. With (54), we compute $\mu_{t|t}^{\text{(p)}}$ in (31) and
with (53), we calculate $(s_{t|t};\ n_{t|t})$ in (40). Finally, with
(53), we find $\mathbb{E}\left\{s_t^{2}\,|\, \mathcal{Y}_t \right\}$,
$\mathbb{E}\left\{s_t n_t\,|\, \mathcal{Y}_t \right\}$ and $\mathbb{E}\left\{n_t^{2}\,|\,y_{t},\, \mathcal{Y}_t \right\}$
in (41). 

\subsection{Summary and discussion of the KF update }

The proposed KF update uses that speech and noise are additive in
the complex STFT domain. The KF update computes the posterior of the
speech and noise log-spectra and of the speech phase. The phase-aware
KF update is presented in (42)-(54) and uses a local prior for the
speech phase from the phase KF prediction in Fig. 3. Using such a
speech phase local prior, which is based on inter-frame speech phase
modelling, is different from using the pitch and intra-frame phase
correlation modelling \cite{a217} \cite{a19}. The KF update computes
the speech phase posterior using the speech phase local prior together
with the speech and noise log-spectra local priors. 

With (54), the complex-valued first trigonometric moment for the speech
phase posterior is estimated. As mentioned in Sec. II.A, the vM and
WN distributions can be computed from the complex-valued first trigonometric
moment with moment matching \cite{a215}. For the phase posterior,
the speech phase concentration, $\kappa_{t|t}^{\text{($\phi$)}}$,
of the vM distribution can be calculated. There is a strong correlation
between $\kappa_{t|t}^{\text{($\phi$)}}$ and the estimated speech
log-spectrum. For voiced frames, $\kappa_{t|t}^{\text{($\phi$)}}$
is large, which means that the posterior variance is small, and the
estimated clean speech log-spectrum, $s_{t|t}$, is high. Conversely,
in silence or unvoiced frames, $\kappa_{t|t}^{\text{($\phi$)}}$ is
small and $s_{t|t}$ is relatively low. 

Figure 4 illustrates the KF prior and posterior distributions when
the KF update presented in (42)-(54) is utilised. Figure 4 depicts
the speech and noise KF posterior and omits the phase KF posterior.
The background in Fig. 4 is based on the covariance matrix of the
KF posterior and the ellipses are the covariance matrices of the KF
prior and posterior distributions. In Fig. 4, we use $y=0$, $\theta=0$
and $\mu^{(\phi)}=0.66$ and $\kappa^{(\phi)}=3.01$ for the vM distribution
of the speech phase, $\phi$. 

In Fig. 4, the speech and noise KF posterior lies within the curvy
triangle, which is mathematically defined by: 

\begin{align}
 & \alpha=1,\ \ \ \ \ \ \ \ \text{cosh}(u)=0.5\exp(-v)\\
 & \alpha=-1,\ \,\ \ \ \ \left|\text{sinh}(u)\right|=0.5\exp(-v)\text{.}
\end{align}

The curvy triangle defines the KF observation constraint region in
the $(s,n)$ plane. The curvy triangle in Fig. 4 and in (55),$\,$(56)
is related to the different values that the phase factor can take
\cite{a225}. In Fig. 4, the KF prior is in the curvy triangle and
on the $u=0$ line, which is for a SNR of $0$ dB. 

Figure 5 resembles Fig. 4 but, now, the KF prior is in the curvy triangle
and off the $u=0$ line. Figures 4,$\,$5 are important because they
show that the KF posterior should lie within the curvy triangle and
the KF observation constraint region. 

\begin{figure}[!t]
\begin{minipage}[t]{0.48\columnwidth}%
\centering \includegraphics[width=0.99\columnwidth]{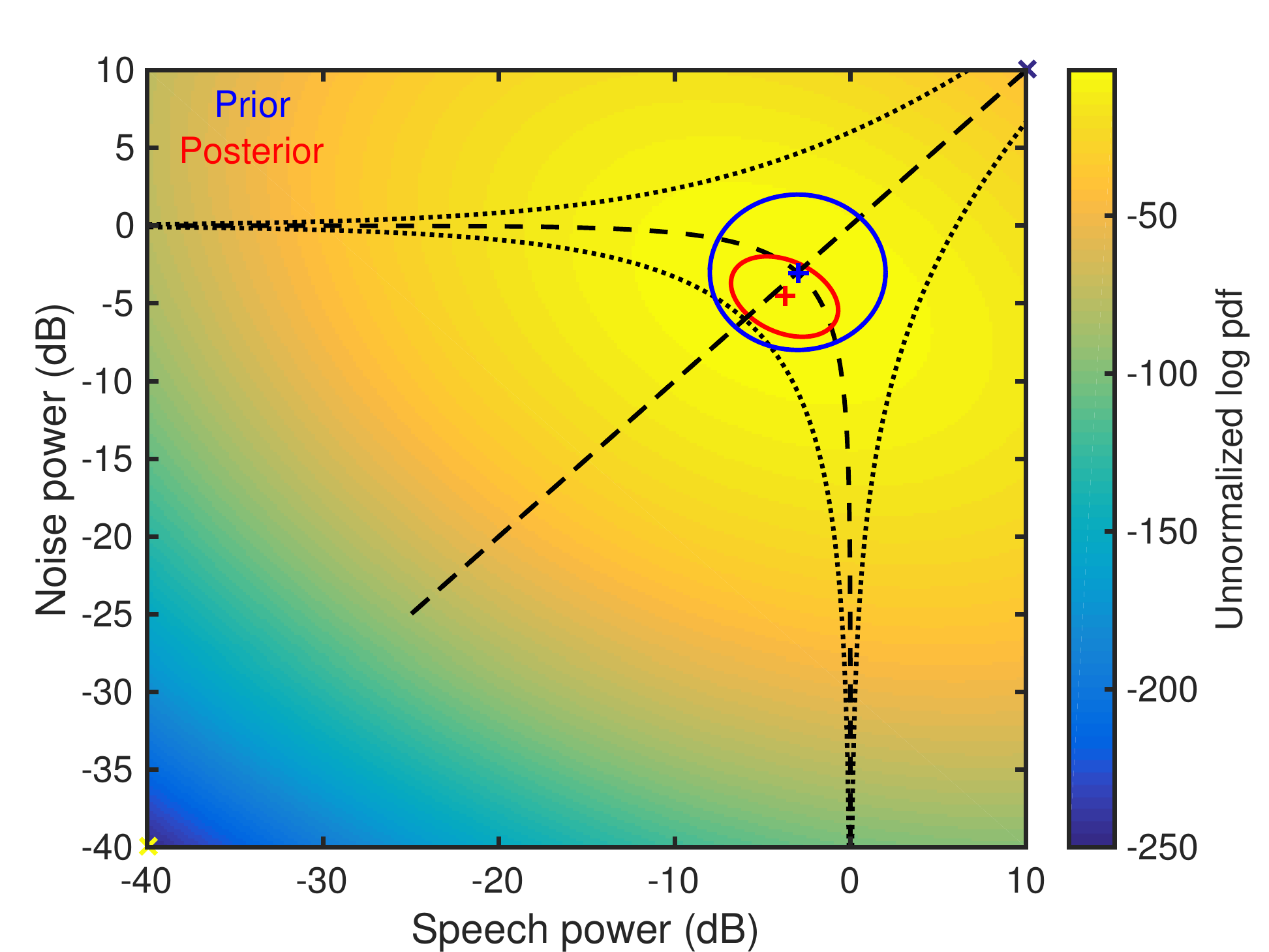}

\caption{Plot of the KF posterior, the KF prior, the $\alpha=-1,0,1$ constraint
lines and the $u=0$ line. }
\end{minipage}\hfill{}%
\begin{minipage}[t]{0.48\columnwidth}%
\centering \includegraphics[width=0.99\columnwidth]{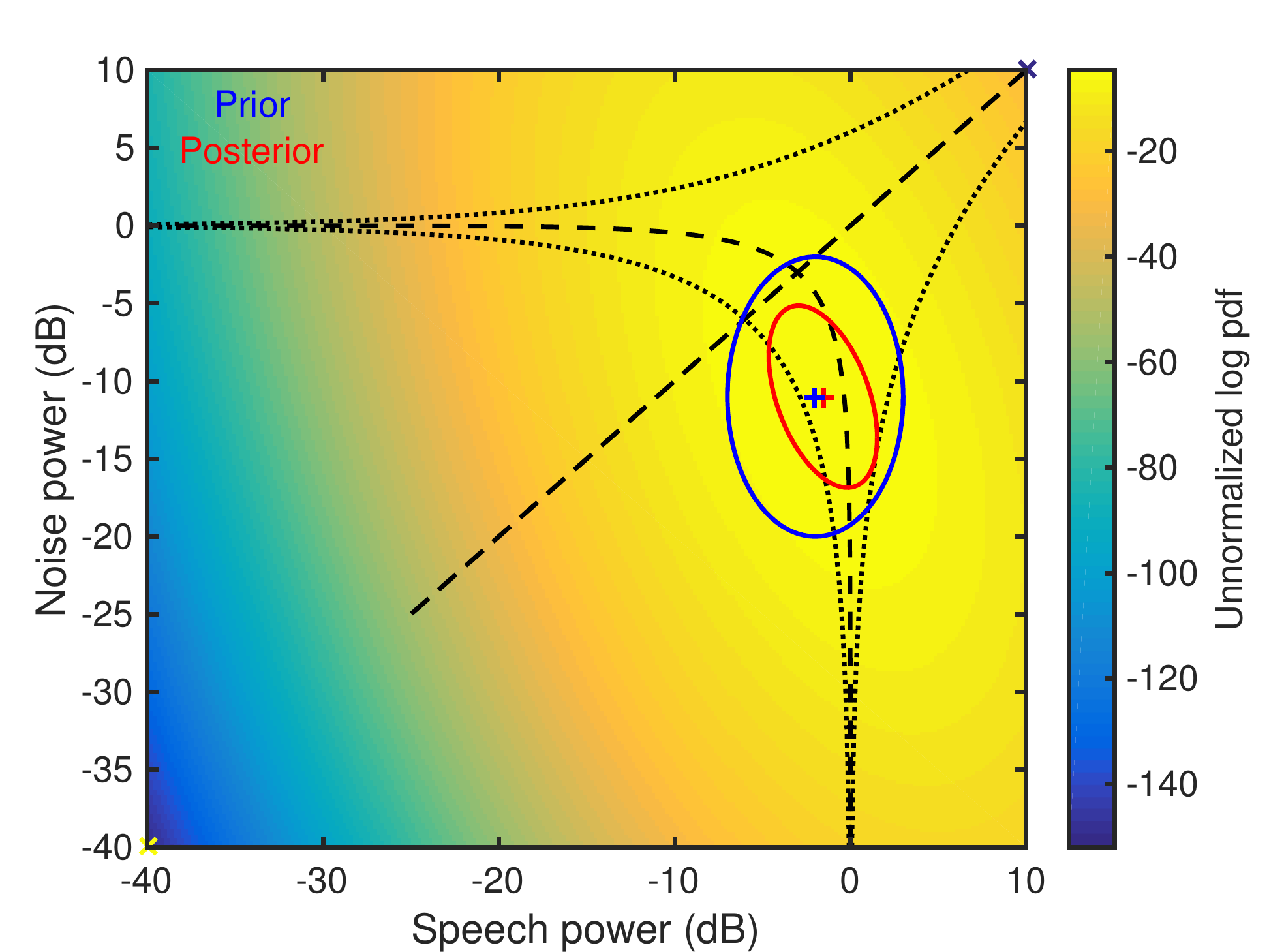}

\caption{Plot of the KF posterior, the KF prior, the $\alpha=-1,0,1$ constraint
lines and the $u=0$ line. }
\end{minipage}
\end{figure}

Finally, equations (42)-(54) describe the proposed nonlinear modified
KF update that is different from the normal linear KF update and is
based on the decorrelation and correlation steps in Sec. III.C. The
KF algorithm uses probability distributions and statistical inference
instead of an observation equation. 

To sum up, in Secs. III.B-D, a Kalman filtering algorithm that performs
$\phi$ tracking using both inter-frame $\phi$ modelling (with the
speech phase KF prediction) and $\phi$ estimation (with the nonlinear
KF update) is designed. The KF is a powerful estimator and its potential
is exploited as far as possible by using speech phase, $\phi$, tracking.
In Sec. III.D, the KF update based on the decorrelation and recorrelation
steps introduced in Sec. III.C is formulated. Finally, as in \cite{a181},
iterated KF updates, and thus two iterations of the proposed nonlinear
KF update step presented in (42)-(54), can be performed. 

\subsection{The non-KF-based priors in the KF loop }

In this section, we present how we use the uncorrelated priors for
speech and noise that are not based on the KF prediction. The non-KF-based
uncorrelated priors for speech and noise are the global speech priors
and the noise priors. 

The KF-based local prior is 

\begin{eqnarray}
 &  & \hspace*{-0.6cm}\hspace*{-0.6cm}p(s_{t},n_{t},\phi_{t}\ |\ \mathcal{Y}_{t-1})=p(s_{t},n_{t}|\mathcal{Y}_{t-1})\ p(\phi_{t}|\mathcal{Y}_{t-1})\text{.}
\end{eqnarray}

In (57), we assume that the speech phase, $\phi$, is independent
of $(s,n)$, as in (49). 

In (57), we note that $p(\phi_{t}|\mathcal{Y}_{t-1})$ is the KF-based
local speech phase prior, the output of the phase KF prediction in
Sec. III.B after using the notation introduced in Sec. II.A and trigonometric
moment matching to the vM distribution. 

The next step is to consider the global speech prior and the noise
prior. Up to this point, we have not included the global speech prior
and the noise prior in our model. If we utilise the global speech
prior and the noise prior, then (57) becomes: 

\begin{eqnarray}
 &  & \hspace*{-0.6cm}\hspace*{-0.6cm}\hspace*{-0.6cm}p\left(s_{t},n_{t},\phi_{t}\ |\ \mathcal{Y}_{t-1},\mathcal{G}^{\text{(s)}},\mathcal{G}^{\text{(n)}}\right)\\
 &  & \hspace*{-0.6cm}\hspace*{-0.6cm}=p\left(s_{t},n_{t}|\mathcal{Y}_{t-1},\mathcal{G}^{\text{(s)}},\mathcal{G}^{\text{(n)}}\right)\ p\left(\phi_{t}|\mathcal{Y}_{t-1}\right)\nonumber 
\end{eqnarray}

where $\mathcal{G}^{\text{(s)}}$ is the global knowledge from the
global speech prior and $\mathcal{G}^{\text{(n)}}$ is the knowledge
from the noise prior. 

In (58), for $p\left(s_{t},n_{t}|\mathcal{Y}_{t-1},\mathcal{G}^{\text{(s)}},\mathcal{G}^{\text{(n)}}\right)$,
we now have: 

\begin{eqnarray}
 &  & \hspace*{-0.6cm}\hspace*{-0.6cm}p\left(s_{t},n_{t}\ |\ \mathcal{Y}_{t-1},\mathcal{G}^{\text{(s)}},\mathcal{G}^{\text{(n)}}\right)\\
 &  & \hspace*{-0.6cm}\propto p\left(s_{t},n_{t}|\mathcal{Y}_{t-1}\right)\ p\left(s_{t}|\mathcal{G}^{\text{(s)}}\right)\ p\left(n_{t}|\mathcal{G}^{\text{(n)}}\right)\text{.}\nonumber 
\end{eqnarray}

In (59), $p\left(s_{t},n_{t}\ |\ \mathcal{Y}_{t-1},\mathcal{G}^{\text{(s)}},\mathcal{G}^{\text{(n)}}\right)$
is the result of the multiplication of the correlated KF-based local
prior for speech and noise, $p(s_t,n_{t}|\mathcal{Y}_{t-1})$, by
the global speech prior and by the noise prior. Both the global speech
prior and the noise prior are assumed to be Gaussian distributions.
When the local prior has a small variance, the algorithm chooses this
estimate because of the Gaussian-Gaussian multiplication in (59). 

\subsection{The processing outside the KF loop }

We now describe the processing that is carried outside the KF loop
(i.e. the dotted rectangle) in Fig. 2. The ``pre-cleaning for speech''
in the ``pre-cleaning for speech or noise'' block in Fig. 2 applies
a conventional enhancement algorithm (e.g. the Log-MMSE estimator
\cite{Ephraim1985}) to the noisy speech, as in \cite{a203} \cite{a233}
and in \cite{a125}. The speech is then divided into overlapping modulation
frames in the log-spectral domain and AR($p$) modelling is performed
to estimate the coefficients $\textbf{a}_t^{\text{(s)}}$, $\zeta_t^{\text{(s)}}$
and $\epsilon_t^{\text{(s)}}$ from (10). AR modelling is performed
using the covariance method, estimating the AR coefficients and the
AR mean. Estimating the speech AR mean, $\zeta_t^{\text{(s)}}$, as
an AR coefficient is beneficial because we operate in the log-spectral
domain and so $\zeta_t^{\text{(s)}}$ represents a scale factor in
the amplitude domain. The AR coefficients need to be estimated from
noisy speech and their estimate is inevitably biased. When speech
pre-cleaning is used, the model misspecification error is reduced
\cite{a203}. 

The ``pre-cleaning for noise'' in the ``pre-cleaning for speech
or noise'' block in Fig. 2 uses a voice activity detection (VAD)
that is based on an AR($1$) one-pole smoother, as in \cite{a123},
and on the estimated SNR from the KF state. The VAD computes $\hat{n}_t = \lambda_t \hat{n}_{t-1} + (1-\lambda_t) y_t$,
where $\hat{n}$ denotes the noise estimate and $\lambda_t$ is obtained
from applying a sigmoid function to the estimated SNR, $\eta_t = \exp(2s_t-2n_t)$.
The estimation of the noise parameters $\textbf{a}_t^{\text{(n)}}$,
$\zeta_t^{\text{(n)}}$ and $\epsilon_t^{\text{(n)}}$ from (15),$\,$(16)
are created from noise AR($q$) modelling in the same way as for speech. 

The ``pre-cleaning for speech phase'' block in Fig. 2 performs speech
phase pre-cleaning using the speech phase estimator in \cite{a19}
prior to the estimate of the speech phase AR model in order to reduce
AR modelling errors. 

In summary, in Sec. III, we presented the KF algorithm in Fig. 2.
We discussed how to track the speech and noise log-spectra in combination
with the speech phase, $\phi$. We also utilised $(s,n,\phi,\psi)$
from which we can define all the other variables presented in Sec.
II including the noisy $y$ and $\theta$. 

\section{Description of the simplified algorithms }

In this section, the various ways in which the algorithm in Sec. III
can be simplified are presented. Simplified algorithms help to identify
which blocks are the most significant. 

\subsection{Tracking only the speech log-spectrum }

We now track speech in the log-spectral domain using one KF for each
frequency bin. Without KF noise tracking, before the KF update step,
the $(s,n)$ priors are uncorrelated. 

We use the speech KF state, $\textbf{x}_t \in \Re^p$, and the speech
KF prediction in (10)-(14). Then, we use the one-element decorrelation
operation, which is a specific case of the multiple-element decorrelation
operation in (35) and (36), before the KF update. The KF update estimates
the posterior distribution of the speech and noise log-spectra. Based
on Sec. II: 

\begin{equation}
e^{2y}=e^{2s}+e^{2n}+2e^{s+n}\alpha\text{.}
\end{equation}

From (60): $\alpha = 0.5\exp\left(2y-s-n\right)-\text{cosh}\left(u\right)$,
where $u=n-s$. It should be noted that $u$ was utilised in Sec.
III.D and in (42). We also use $y = 0.5(s+n+\log(2(\alpha+\cosh(u))))$. 

The variable transformation $(s,n,\alpha) \Rightarrow (u,y,\alpha)$
is then performed. The Jacobian determinant is $\Delta_{2}=1$. Now,
the posterior distribution given the noisy log-amplitude, $y$, is:
\begin{align}
 & p(u,\alpha\,|\,y)=\left.\frac{p(u,\alpha,\,y)}{p(y)}\right|_{y}\propto\left.\left(p(s,n)\ p(\alpha)\left|\Delta_{2}\right|^{-1}\right)\right|_{y}\\
 & \propto p(\alpha)\ \mathscr{N}\left(\left(\begin{array}{c}
2y-\log\left(2(\alpha+\cosh(u))\right)-u\\
2y-\log\left(2(\alpha+\cosh(u))\right)+u
\end{array}\right);\,\mathbf{m},\,\mathbf{S}\right)\nonumber 
\end{align}

where a Gaussian distribution with mean $\mathbf{m}$ and covariance
matrix $\mathbf{S}$ is used for $p(s,n)$, as in \cite{a225}. We
find the moments of the posterior distribution using 

\begin{align}
 & \mathbb{E}\{s^{a}n^{b}\ |\ y\}=\int_{\alpha=-1}^{1}\int_{u=-\infty}^{\infty}s^{a}n^{b}\ p(u,\alpha\ |\ y)\ du\ d\alpha\nonumber \\
 & =\dfrac{1}{|\Delta_{2}|\ p(y)}\int_{\alpha=-1}^{1}p(\alpha)\int_{u=-\infty}^{\infty}s^{a}n^{b}\ p(s,n)\ du\ d\alpha
\end{align}

where $0 \leq a+b \leq 2$, $s=s(u,y,\alpha)$ and $n=n(u,y,\alpha)$. 

We update the mean and covariance matrix of the current speech-noise
KF state. We calculate the first and second moments of the posterior
distribution of the current speech and noise log-spectra, $(s,n)$,
using $E \{ s^a n^b | y \}$ in (60),$\,$(61). 

\subsection{Alternative KF updates }

The KF updates in Sec. III.D and Sec. IV.A use that speech and noise
are additive in the complex STFT domain. Four alternative ways to
do the KF update are now considered. The first way is to assume speech
and noise additivity in the power spectral domain and that speech
and noise follow a Gaussian distribution in the log-spectral domain.
We denote this method by AP that refers to additivity in the power
spectral domain. Assuming additivity in the power spectral domain,
which is equivalent to assuming $\alpha=0$ and that speech and noise
are in-quadrature, is an intermediate case of the proposed KF updates
in Sec. III.D and Sec. IV.A that use $\alpha \in [-1,1]$. 

The second alternative way is to assume speech and noise additivity
in the amplitude domain and that speech and noise are Gaussian in
the log-spectral domain. We denote this technique by AA that refers
to additivity in the amplitude spectral domain. Assuming speech and
noise additivity in the amplitude domain, which is equivalent to assuming
$\alpha=1$ and that speech and noise are in-phase, is an extreme
case of the proposed KF updates in Sec. III.D and Sec. IV.A. 

The third alternative way is to assume speech and noise additivity
in the power spectral domain and that the speech and noise are Gaussian
in the power domain. To do this, we convert the log-power domain speech
and noise distributions into the power domain by matching the first
and second moments. We denote this method by APPG that refers to additivity
in the power spectral domain with power-domain Gaussians. 

The fourth alternative way is to assume speech and noise additivity
in the amplitude domain and that the speech and noise are Gaussian
in the amplitude domain. To do this, we convert the log-domain speech
and noise distributions into the amplitude domain by matching the
first and second moments. We denote this technique by AAAG that refers
to additivity in the amplitude domain with amplitude-domain Gaussians. 

\subsection{Tracking only the speech and noise log-spectra }

We now describe a simplified version of the KF algorithm of Fig. 2
that tracks the speech and noise log-spectra. Based on III.B, with
noise tracking, the KF state $\in \Re^{p+q}$ is the speech KF state
$\in \Re^{p}$ and the noise KF state $\in \Re^{q}$. Moreover, with
noise tracking, the $(s,n)$ local priors are correlated: we track
the correlation between the speech and noise log-spectra. 

We use the joint speech-noise KF state, $\textbf{x}_t^{(j)}$, and
the speech and noise KF prediction described in Sec. III.B and in
(10)-(16). We then use the multiple-element decorrelation operation
in (35) and (36), before the KF update. Next, we perform the KF update
that is presented in Sec. IV.A and in (61)-(62). 

We now define the algorithm acronyms for the proposed Kalman filtering
algorithms. We denote the algorithm presented in Sec. IV.A by ST that
refers to speech tracking (ST). Correspondingly, we denote the KF
algorithm presented in Sec. IV.C by SNT that refers to ST and noise
tracking (NT). In addition, we also denote the Kalman filtering algorithm
that performs speech phase estimation presented in Sec. III by SNPT
that refers to SNT and phase tracking (PT). 

\section{Implementation }

For the implementation of the Kalman filtering algorithm, we use acoustic
frames of length $32$ ms, an acoustic frame time increment of $8$
ms, modulation frames of $64$ ms and a modulation frame time increment
of $8$ ms. We use a full overlap for the modulation frames. We use
speech utterances from the TIMIT database \cite{a187} sampled at
$16$ kHz. We use noise recordings from the RSG-10 database \cite{a58}
at SNR levels from $-20$ dB to $30$ dB. Randomly selected noise
segments are used in each test and noisy speech is created using \cite{a59}. 

For the noise prior, we use external noise estimation based on \cite{a76}.
For speech amplitude spectrum pre-cleaning, we use the Log-MMSE estimator
\cite{Ephraim1985}. The KF state dimensions for speech and noise
are respectively $p=2$ and $q=2$. In addition, for speech phase
pre-cleaning in Fig. 2, we utilise \cite{a19}. 

\subsection{Evaluation of Integrals }

We now present the integration details of the KF update of the proposed
algorithm in Sec. III.D and in Sec. IV.A. In (52)-(54) and in (62),
we integrate over the phase factor and over $u$. The outer integration
over the phase factor, $\alpha$, is done with $R$ sigma points \cite{a225}.
The inner integration over $u$ is performed either by numerical integration
or by approximating the range in segments with truncated Gaussian
distributions. 

We utilise $\displaystyle \mathbb{E}\{\alpha^{c}\}$, and $\displaystyle \mathbb{E}\{\alpha^{c}\}=2^{-c} c! \left(\left(0.5c\right)!\right)^{-2}$
for even $c$ and zero otherwise, for the outer integration over $\alpha$
in (52)-(54) and in (62). The outer integration over $\alpha$ is
done using the Unscented transform \cite{a154} and weighted sigma
points for approximating integration with summation \cite{a109}.
We fit the first six moments of the phase factor, $\alpha$, using
(1) and 

\begin{align}
\mathbb{E}\{\alpha^{c}\}=\int_{\alpha=-\infty}^{\infty}\ p(\alpha)\ \alpha^{c}\ d\alpha=\sum_{r=1}^{R}w_{r}\ \alpha_{r}^{c}\text{.}
\end{align}

The outer integral over $\alpha$ in (52)-(54) and in (62) is exact
to $f(\alpha)$ polynomials up to the fifth order in (63), when $R=3$.
For $R=3$, $w_r=0.333 \ \forall r$ and $\alpha_r = \{ 0, \pm \sqrt{0.75} \}$.
Using $R=3$ sigma points produces better results than using $R=1$
and using $R=5$ produces similar results to using $R=3$. 

For the integration over $u$, we use numerical integration. An alternative
way is to utilise straight line segments with truncated Gaussian distributions
in order to obtain a closed-form approximation. Based on Sec. III.E
and on Figs. 4-5 that show the $(s,n)$ plane, for $v$ in terms of
$u$, it is possible to approximate the range in segments with truncated
Gaussian distributions. In (52)-(54) and in (62), a small number of
straight line segments in which $v$ is linearly approximated in $u$
for a given $\alpha$ can be used. In (52)-(54), $\phi$ is a complicated
function of $u$ and a small number of straight line segments in which
$\phi$ is linearly approximated in $u$ for a given $\alpha$ can
also be utilised. The function of $\phi$ in terms of $u$ is in (45)-(48). 

\subsection{Computational complexity }

In this section, we provide a brief analysis on the computational
requirements of the different configurations of the proposed modulation-domain
Kalman filtering algorithm. The estimation formulas in Secs. III-IV
describe the computation for each frame of noisy speech and for each
frequency bin. 

As in \cite{a179}, we compute the average real-time factor so as
to evaluate the computational efficiency (i.e. the latency) of the
presented algorithm. For the simplified KF algorithm in Sec. IV.A
with an acoustic frame length of $32$ ms, an acoustic time increment
of $16$ ms, a modulation frame of $128$ ms and a modulation time
increment of $64$ ms, the average real-time factor is $28$ using
a laptop equipped with an Intel Core i5 processor. Using the same
machine, the average real-time factor of the algorithm in \cite{a179}
is $4$; the algorithm in \cite{a179} is $7$ times faster than the
proposed algorithm leading to the conclusion that one of the main
disadvantages of the proposed Kalman filtering algorithm is its computational
efficiency. 

For the algorithm in Sec. III, the real-time factor is between $30$
and $50$ using the same machine, dependent on the acoustic and modulation
frame parameters. In Sec. III, we reconstruct the speech phase for
all frequencies and not up to $4$ kHz, as in \cite{a19}, because
the nonlinear KF update in Sec. III. D is a joint speech log-amplitude
and phase estimator and, in high frequencies, estimating the speech
amplitude is important. 

We observe a tradeoff between accuracy and computational complexity.
For high accuracy, we want a small acoustic frame time increment \cite{a11}
and a short modulation frame length. 

For a perceptual comparison, the reader is referred to \cite{a244}
where some recordings processed by the KF are available. 

\section{Evaluation }

The KF algorithm is compared with certain state-of-the-art approaches
to speech enhancement. We compare the different algorithm configurations
in Secs. III-IV with: (a) the traditional technique of Log-MMSE \cite{Ephraim1985}
with the MMSE noise estimator \cite{a76}, and (b) the optimally modified
log-spectral amplitude (OMLSA) estimator \cite{Cohen2001} with the
improved minima controlled recursive averaging (IMCRA) noise estimator
\cite{Cohen2002}. 

We compare the ST algorithm with the alternative KF updates that are
introduced and discussed in Sec. IV.B. The alternative KF updates
in Sec. IV.B constitute the KF baselines of this paper and the algorithms
(a)-(b), in the previous paragraph, constitute the non-KF baselines
of this paper. 

Table I summarizes the nine different algorithms that are considered
in this section. ST, SNT and AP were also evaluated in \cite{a225}
and ST and SNT were also evaluated in \cite{a236}. Based on Sec.
IV.B, Table I presents the different combinations of the signal model
and the KF state. The horizontal bold labels correspond to the KF
state domains that define the domain in which the KF prediction is
performed. In addition, the vertical bold labels correspond to the
signal model domains that define the domain in which additivity between
speech and noise is assumed and/or used in the KF update step. 

\ctable[caption = \textrm{\normalfont The $9$ different algorithms that are considered and the different combinations of the domains of the signal model and the KF state. The rows and columns refer to the signal model domains and the KF state domains respectively}.,   
label = tab:fpsaaaa6a324234asafaada55aasf22232ahhaaaaaaaafafiiuuuaaaga3a33aaa0a0a0aaa0aaaaaooiiiaiaiaaaaaaaaaaaa,   
pos = hp,   
doinside=\hspace*{0pt}, width=.49\textwidth ]{p{2.1cm} p{2.0cm} p{1.8cm} p{1.7cm}}{}{ 
\FL & \textbf{Log-spectral} & \textbf{Amplitude} & \textbf{Power} 
\NN \textbf{STFT} & ST & - & - 
\NN \textbf{Amplitude} & AA & AAAG & - 
\NN \textbf{Power} & AP & - & APPG 
\ML SNT & SNPT & Log-MMSE & OMLSA 
\LL } 

The structure of the evaluation section is as follows. Section VI.A
presents the evaluation metrics and Sec. VI.B compares the Kalman
filtering algorithm with the baselines Log-MMSE and OMLSA. Section
VI.C examines the signal model and KF state domains and Sec. VI.D
examines the quantities that are tracked. Based on Secs. I-IV and
on Table I, Kalman filtering depends on the signal model domain, on
the KF state domain and on the quantities that are tracked. Finally,
Sec. VI.E considers the quality metrics presented in Sec. VI.A. 

\subsection{Quality metrics }

The algorithms are evaluated in terms of speech quality with the Perceptual
Evaluation of Speech Quality (PESQ) metric \cite{ITU_T_P862} and
with the speech quality (SIG), the noise quality (BAK) and the overall
speech quality (OVRL) metrics \cite{a205}. 

\subsection{Comparison with the non-KF baseline algorithms }

In this section, we evaluate the different configurations of the proposed
modulation-domain Kalman filtering algorithm. We use the core test
set of the TIMIT database in order to evaluate the presented algorithm
and its different configurations in different noise types. The number
of sentences in the TIMIT core test is $192$; the TIMIT core test
contains $16$ male and $8$ female speakers each reading $8$ distinct
sentences. The algorithms are examined across different speaking rates. 

\begin{figure}[t]
\begin{minipage}[t]{0.48\columnwidth}%
\centering\includegraphics[width=0.99\columnwidth]{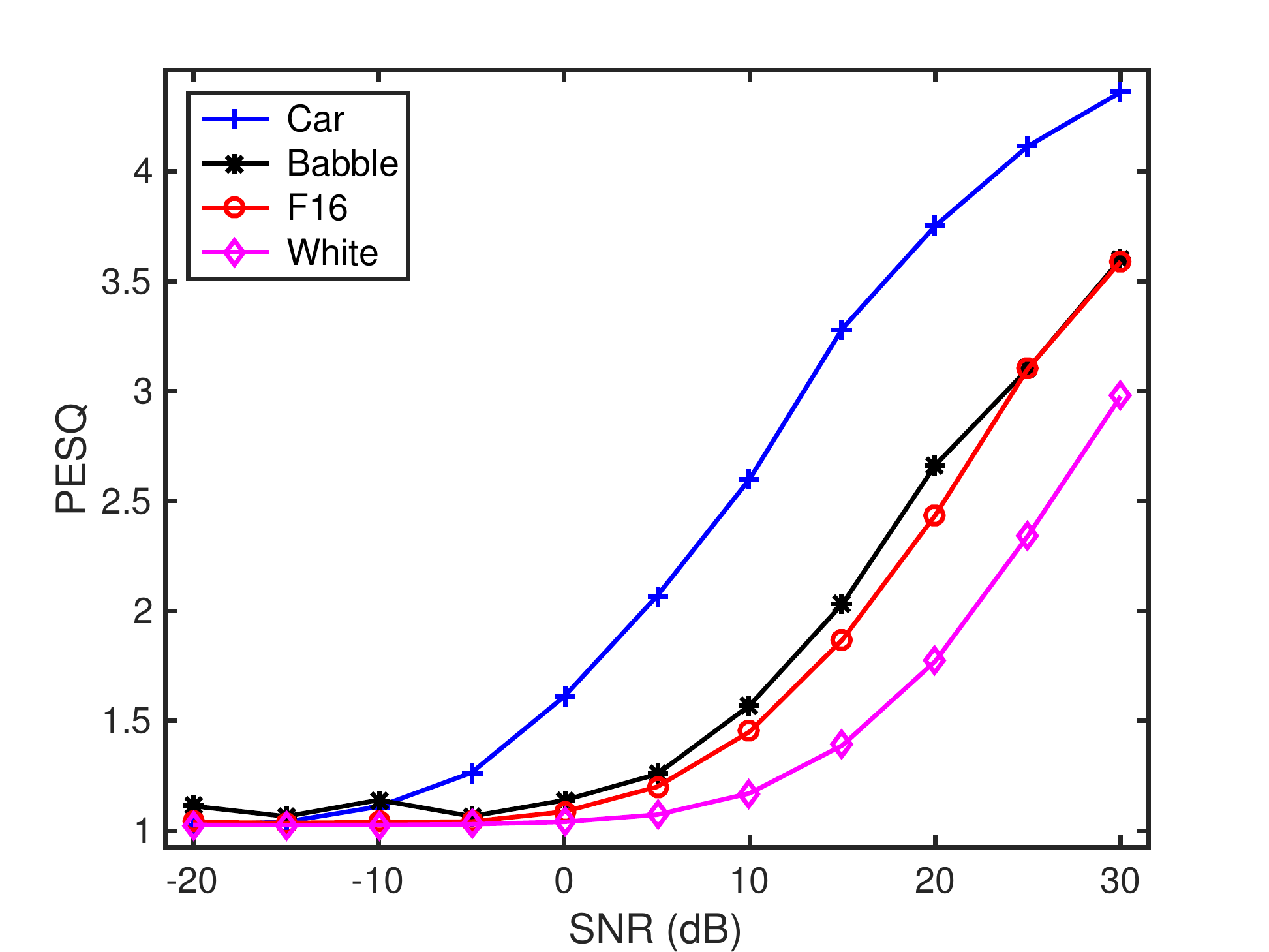}

\caption{Plot of the mean PESQ scores for noisy speech versus SNR for the noise
types used in evaluation. }
\end{minipage}\hfill{}%
\begin{minipage}[t]{0.48\columnwidth}%
\centering \includegraphics[width=0.99\columnwidth]{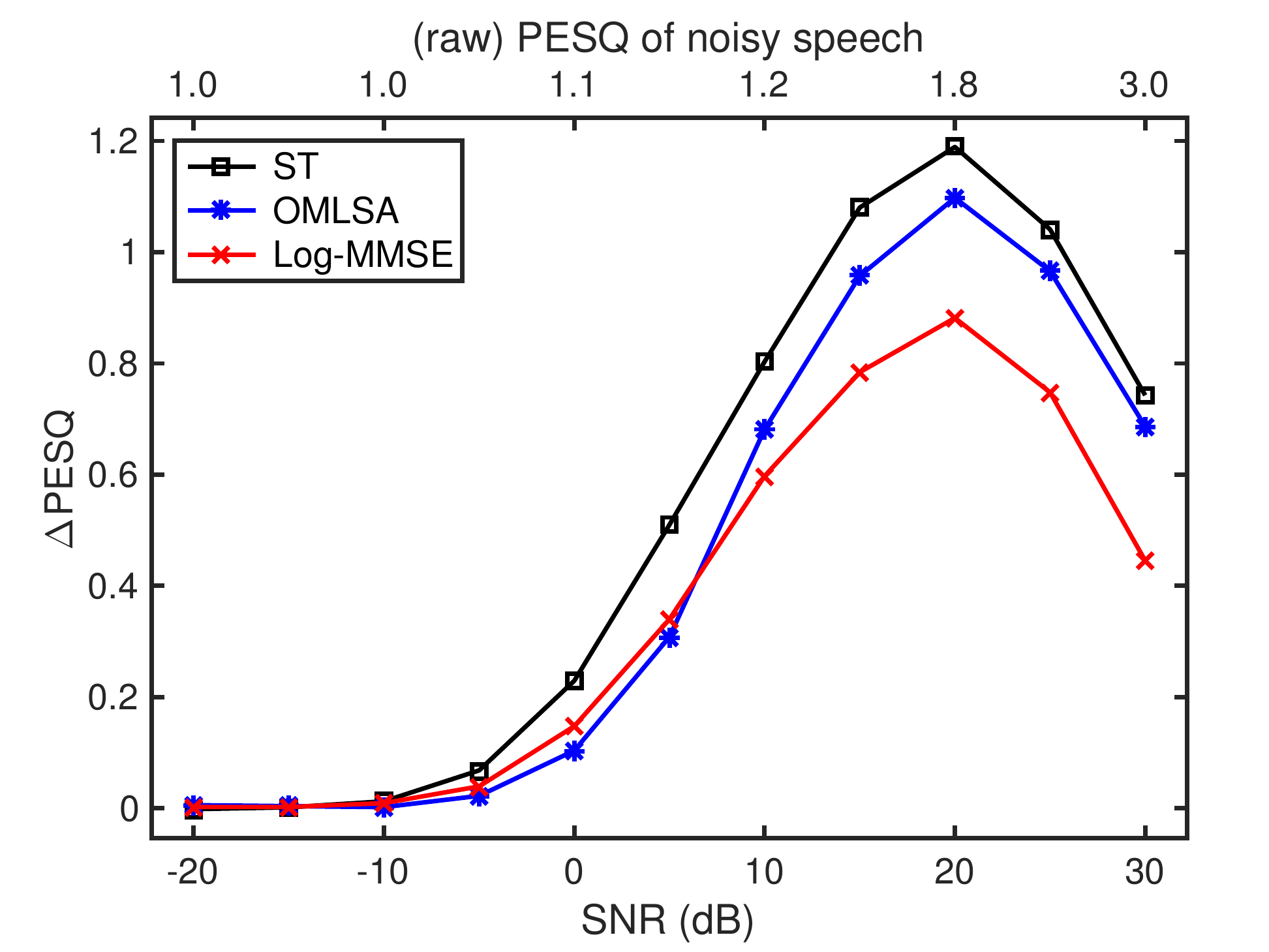}

\caption{Plot of the mean $\Delta$PESQ for speech with white noise versus
SNR for the speech-only KF (ST) and two conventional algorithms. }
\end{minipage}
\end{figure}

\begin{figure}[t]
\begin{minipage}[t]{0.48\columnwidth}%
\centering \includegraphics[width=0.98\columnwidth]{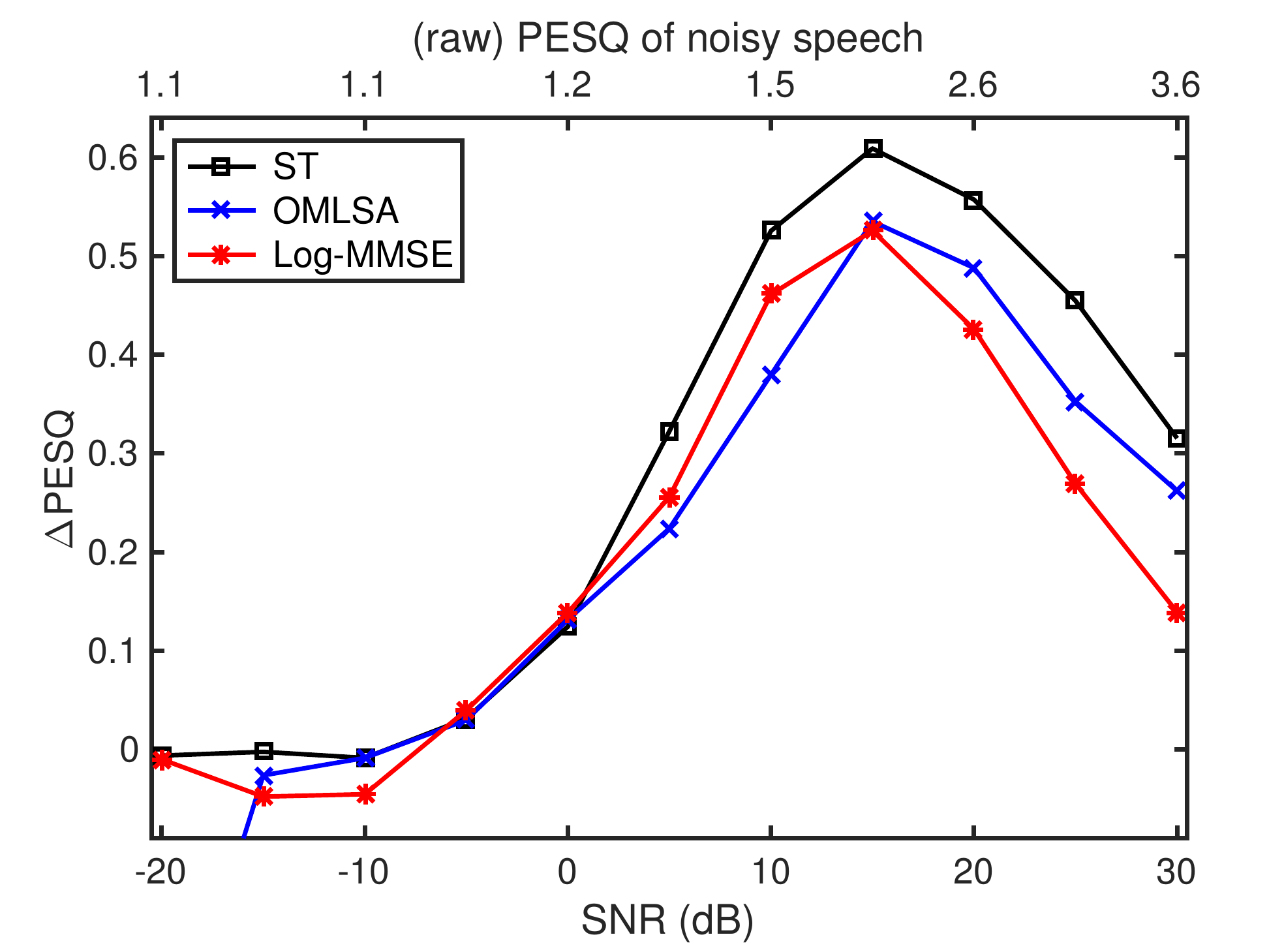}

\caption{Plot of the mean $\Delta$PESQ for speech with additive babble noise. }
\end{minipage}\hfill{}%
\begin{minipage}[t]{0.48\columnwidth}%
\centering \includegraphics[width=0.98\columnwidth]{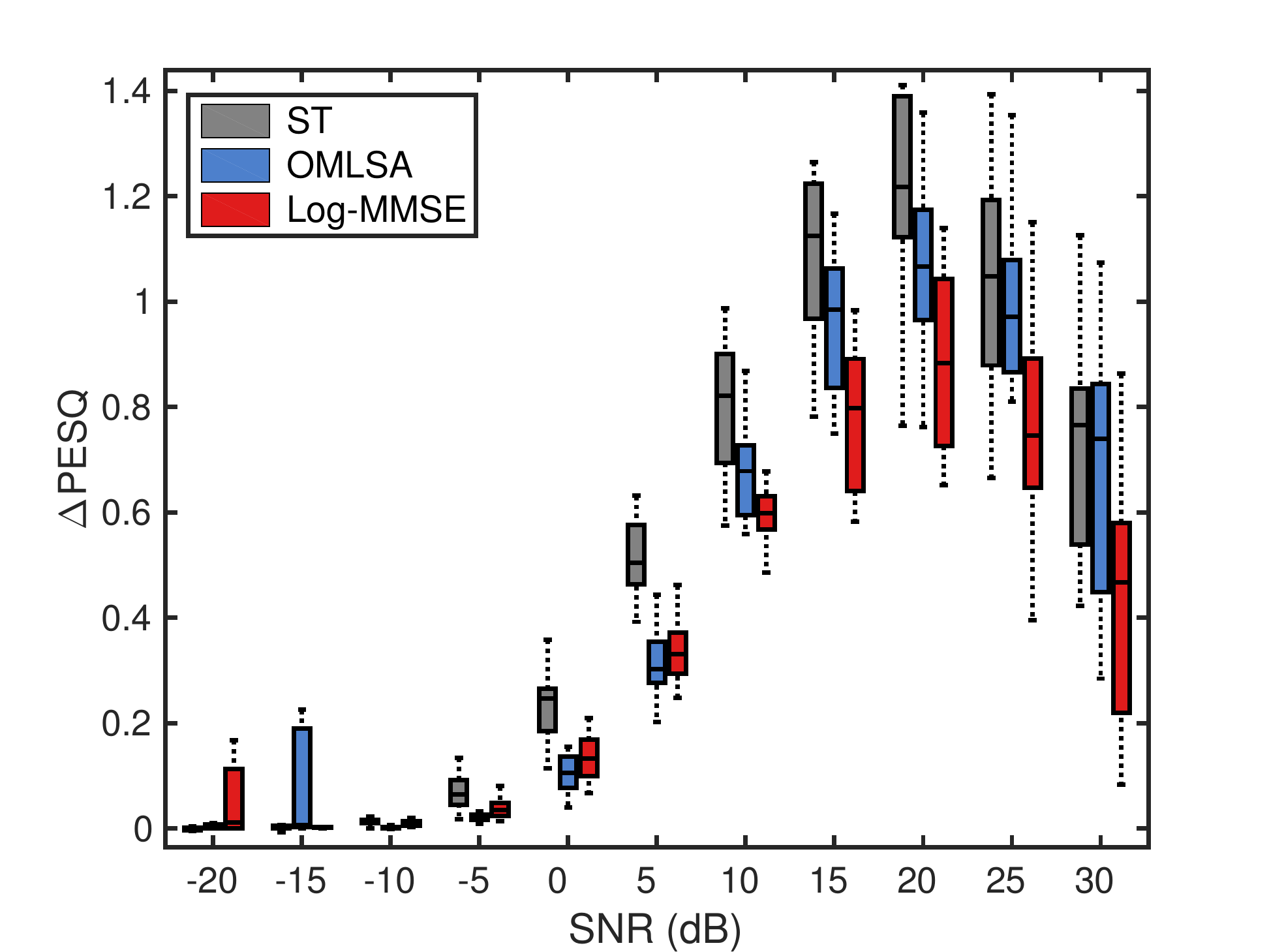}\caption{Boxplot of the $\Delta$PESQ for speech with additive white noise. }
\end{minipage}
\end{figure}

Figure 6 shows the different types of noise that are used in Sec.
VI for evaluation purposes and presents the (raw) PESQ scores for
the noisy speech signals. Figure 6 depicts the raw PESQ scores for
car, babble, F16 aircraft and white noises. As seen in Fig. 6, the
noise types of babble and F16 have small differences in their effects
at a particular SNR. 

The main enhancement algorithm of this paper is SNPT, which is presented
in Sec. III. We evaluate SNPT and we also evaluate the simpler algorithms
of ST and SNT, which are presented in Sec. IV, so as to assess the
relative contribution of each of the blocks of the main algorithm.
In this way, the contribution of ST and SNT to the final PESQ improvement,
$\Delta$PESQ, of SNPT will be assessed. The ST algorithm is first
evaluated and compared with the two non-KF baselines, Log-MMSE and
OMLSA, using $\Delta$PESQ. Then, the more complex versions of the
Kalman filtering algorithm, SNT and SNPT, will be compared with ST
using $\Delta$PESQ. 

Figure 7 illustrates the mean of the $\Delta$PESQ scores of ST, OMLSA
and Log-MMSE for white noise in various SNRs. The ST algorithm is
better than the other algorithms for $-10 < \text{SNR} \leq 30$ dB.
In Fig. 7, the top horizontal axis shows the (raw) PESQ of noisy speech
for white noise; the top horizontal axis in Fig. 7 matches the white
noise curve in Fig. 6. Figure 6 shows that white noise decreases PESQ
more than the other noise types of car, babble and F16. Moreover,
Fig. 7 shows that the enhancement algorithms increase PESQ a lot because
white noise is a stationary non-speech-shaped noise type. 

Figure 8 depicts the mean $\Delta$PESQ scores of ST, OMLSA and Log-MMSE
for babble noise. We observe that the $\Delta$PESQ scores are lower
in Fig. 8 than in Fig. 7 since babble noise is non-stationary. Figure
8 shows that ST is better than the two non-KF baseline algorithms
for $0 < \text{SNR} \leq 30$ dB. In Fig. 8, the top horizontal axis
shows the (raw) PESQ of noisy speech for babble noise; in other words,
the top horizontal axis in Fig. 8 matches the black babble noise curve
in Fig. 6. 

Figure 9 illustrates the median and the interquartile range of the
$\Delta$PESQ scores of ST, OMLSA and Log-MMSE for white noise. Based
on both Figs. 7 and 9, the ST algorithm is better than the two baseline
algorithms for $-10 < \text{SNR} \leq 30$ dB. 

The legends of all the figures, and of all the boxplots like Fig.
9, have the same order as the curves in high SNRs. 

Figure 10 shows the median and the interquartile range of the $\Delta$PESQ
scores of ST, OMLSA and Log-MMSE for babble noise. Likewise, Fig.
11 illustrates the median and the interquartile range of the $\Delta$PESQ
scores of ST, OMLSA and Log-MMSE for F16 noise. Figure 10 shows that
ST is better than the two non-KF baselines for $5 < \text{SNR} \leq 30$
dB for babble noise. In addition, Fig. 11 shows that ST is better
than the two baselines for $0 < \text{SNR} \leq 30$ dB for F16 noise. 

Figures 9-11 are boxplots and examine the variability and the inter-quartile
range of the $\Delta$PESQ scores. We observe that the three algorithms
do not have comparable $\Delta$PESQ variability. Boxplots also illustrate
the frequent best and worst enhancement cases; for clarity, the infrequent
best and worst enhancement cases (i.e. the outliers) are not included.
Based on Fig. 9, at $20$ dB SNR white noise, ST can have a $\Delta$PESQ
of $1.41$ in the frequent case and a $\Delta$PESQ of $0.79$ in
the frequent worst case. Also, based on Fig. 10, at $15$ dB SNR babble,
ST can have a $\Delta$PESQ of $0.74$ in the frequent best case and
a $\Delta$PESQ of $0.49$ in the frequent worst case. 

\begin{figure}[t]
\begin{minipage}[t]{0.48\columnwidth}%
\centering \includegraphics[width=0.96\columnwidth]{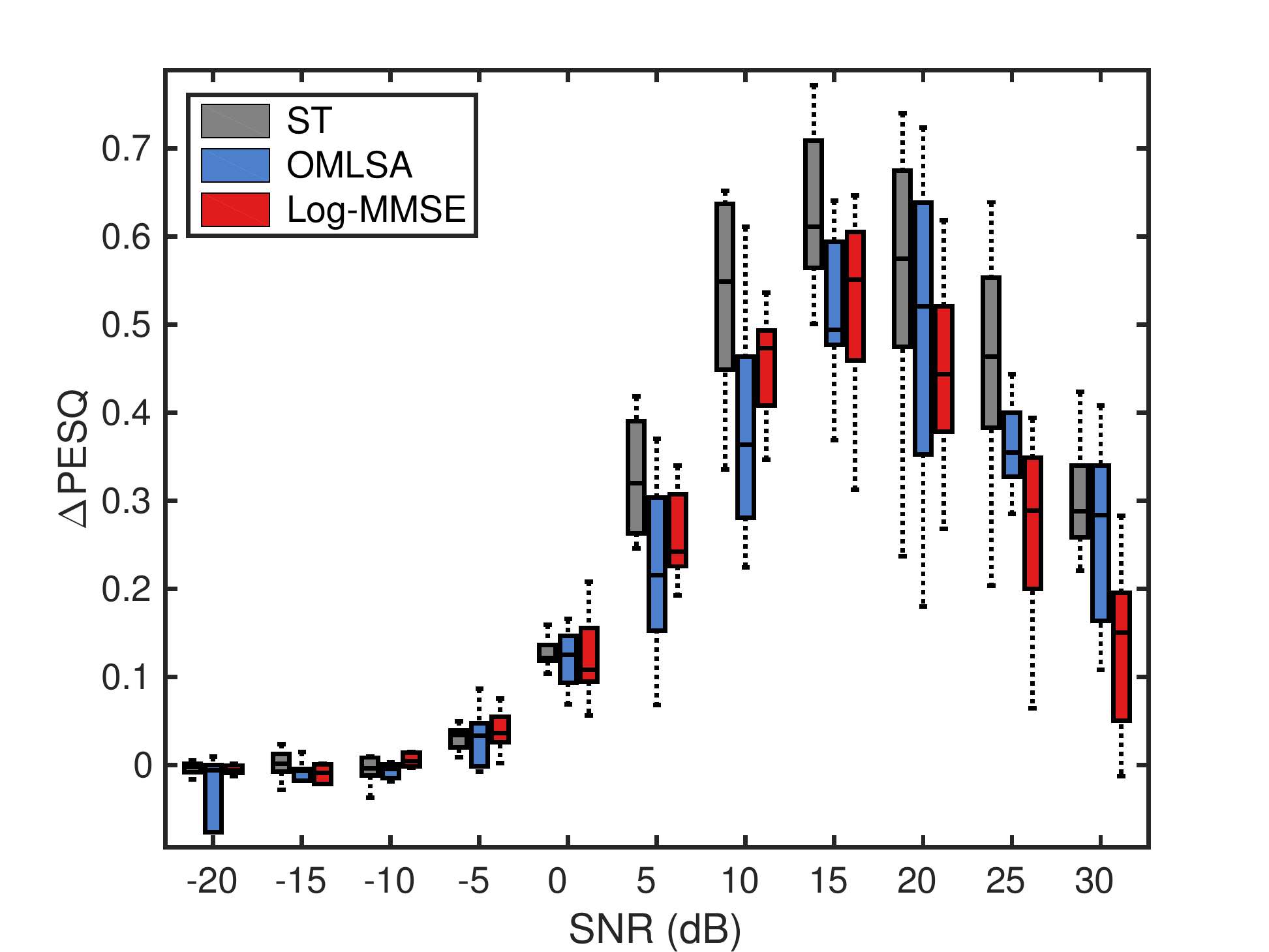}\caption{Boxplot of the $\Delta$PESQ for speech with additive babble noise. }
\end{minipage}\hfill{}%
\begin{minipage}[t]{0.48\columnwidth}%
\centering \includegraphics[width=0.96\columnwidth]{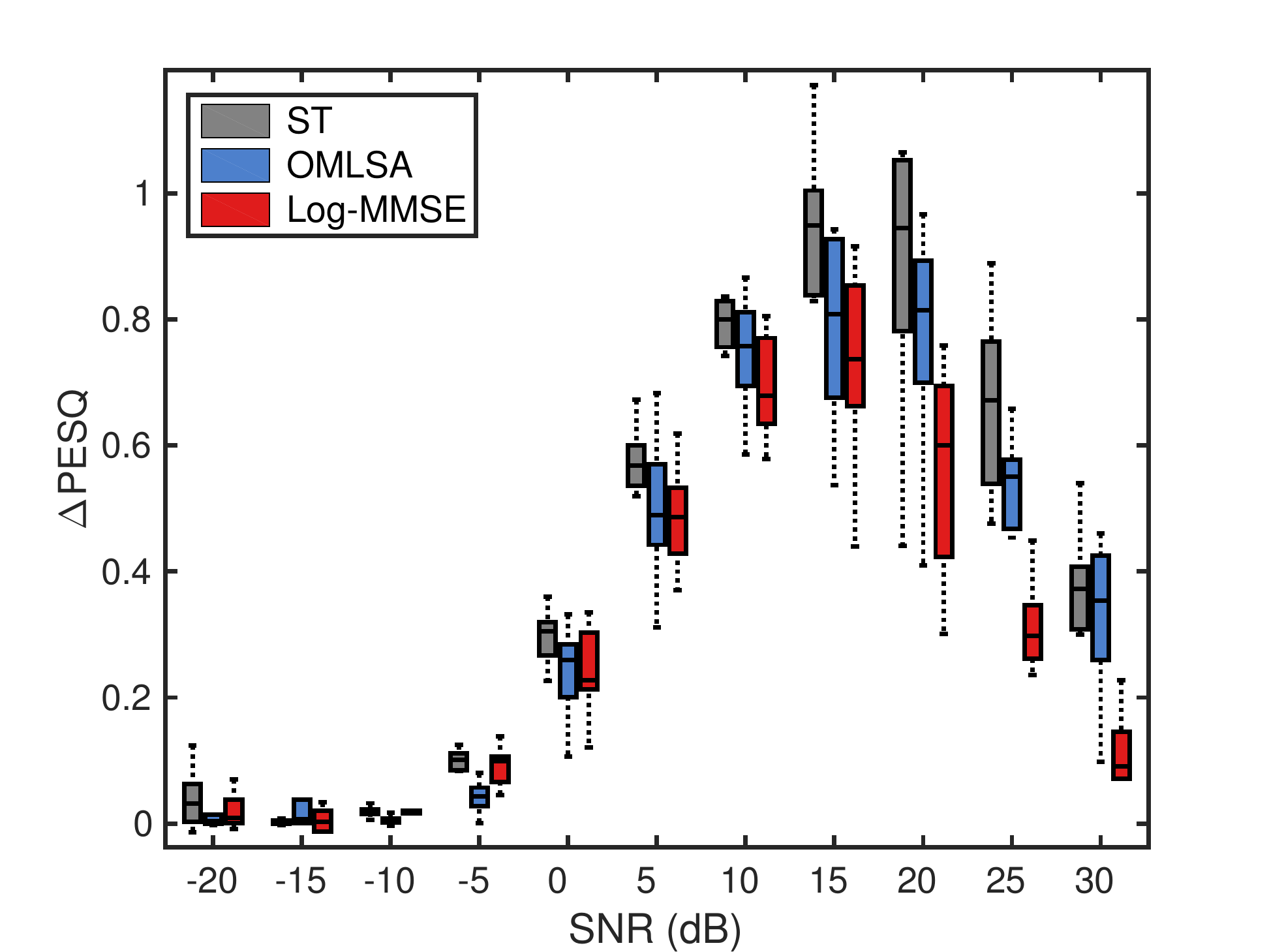}\caption{Boxplot of the $\Delta$PESQ scores for speech with F16 noise. }
\end{minipage}
\end{figure}

Figure 12 is a boxplot of the $\Delta$PESQ scores of ST, Log-MMSE
and OMLSA for car noise. Figure 12 shows that ST is better than and/or
comparable to the Log-MMSE and OMLSA baselines for the SNR range of
$-10 < \text{SNR} \leq 30$ dB. 

\begin{figure}[t]
\begin{minipage}[t]{0.48\columnwidth}%
\centering \includegraphics[width=0.96\columnwidth]{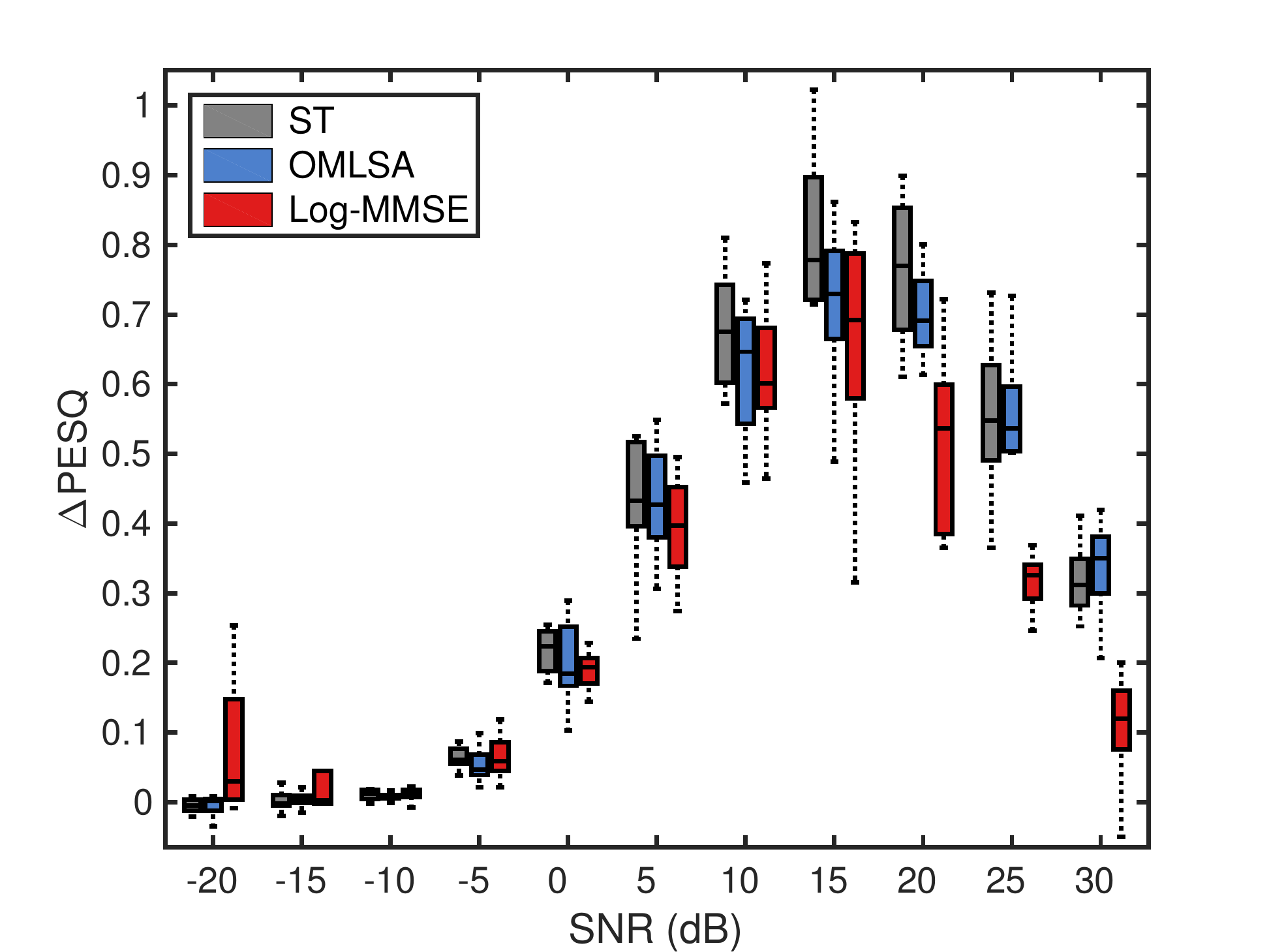}\caption{Boxplot of the $\Delta$PESQ for speech with additive car noise. }
\end{minipage}\hfill{}%
\begin{minipage}[t]{0.48\columnwidth}%
\centering\includegraphics[width=0.96\columnwidth]{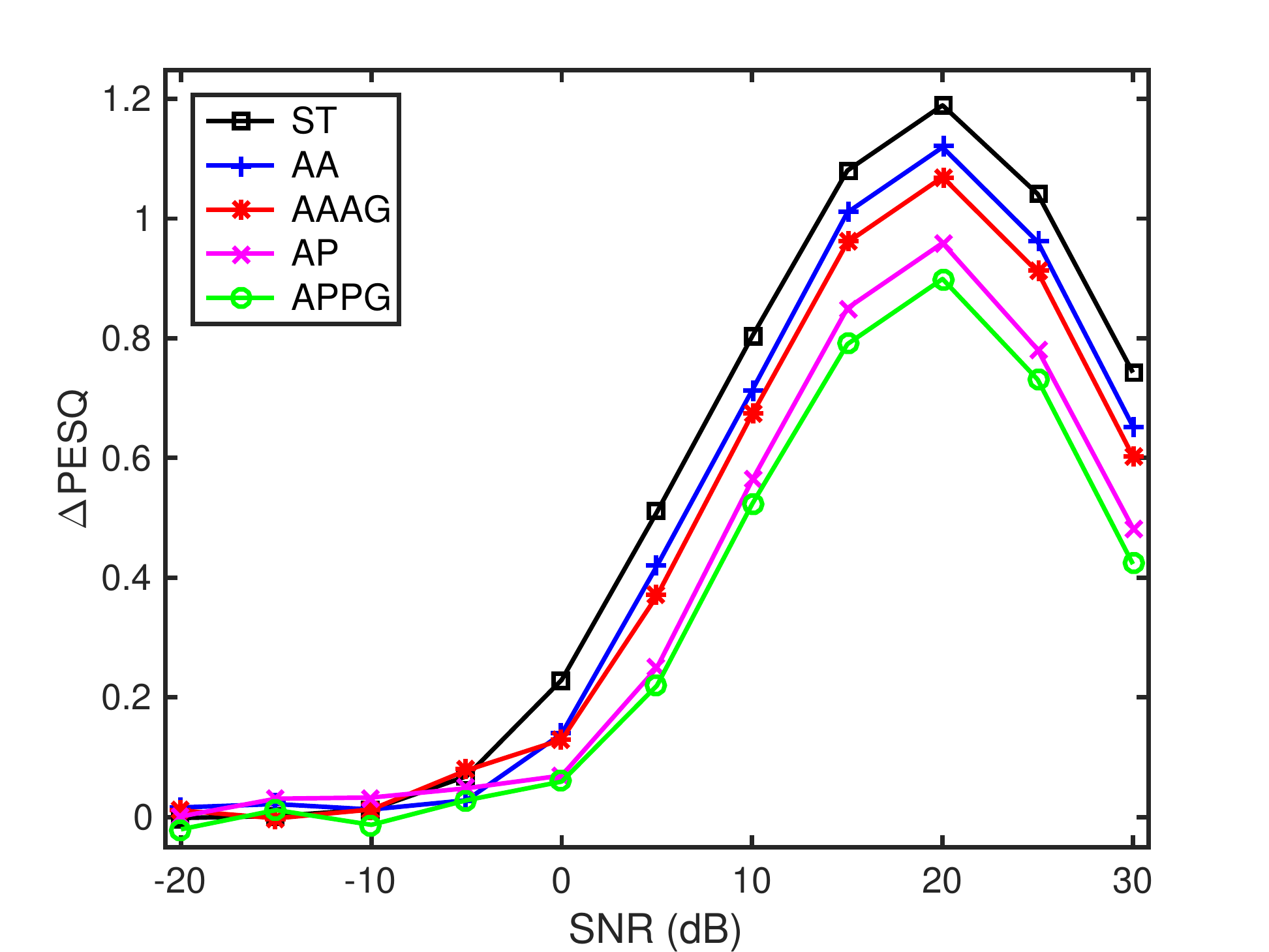}\caption{Plot of the mean $\Delta$PESQ for speech with additive white noise
versus SNR showing the effect of using different signal models. }
\end{minipage}
\end{figure}

\begin{figure}[t]
\begin{minipage}[t]{0.48\columnwidth}%
\centering \includegraphics[width=0.96\columnwidth]{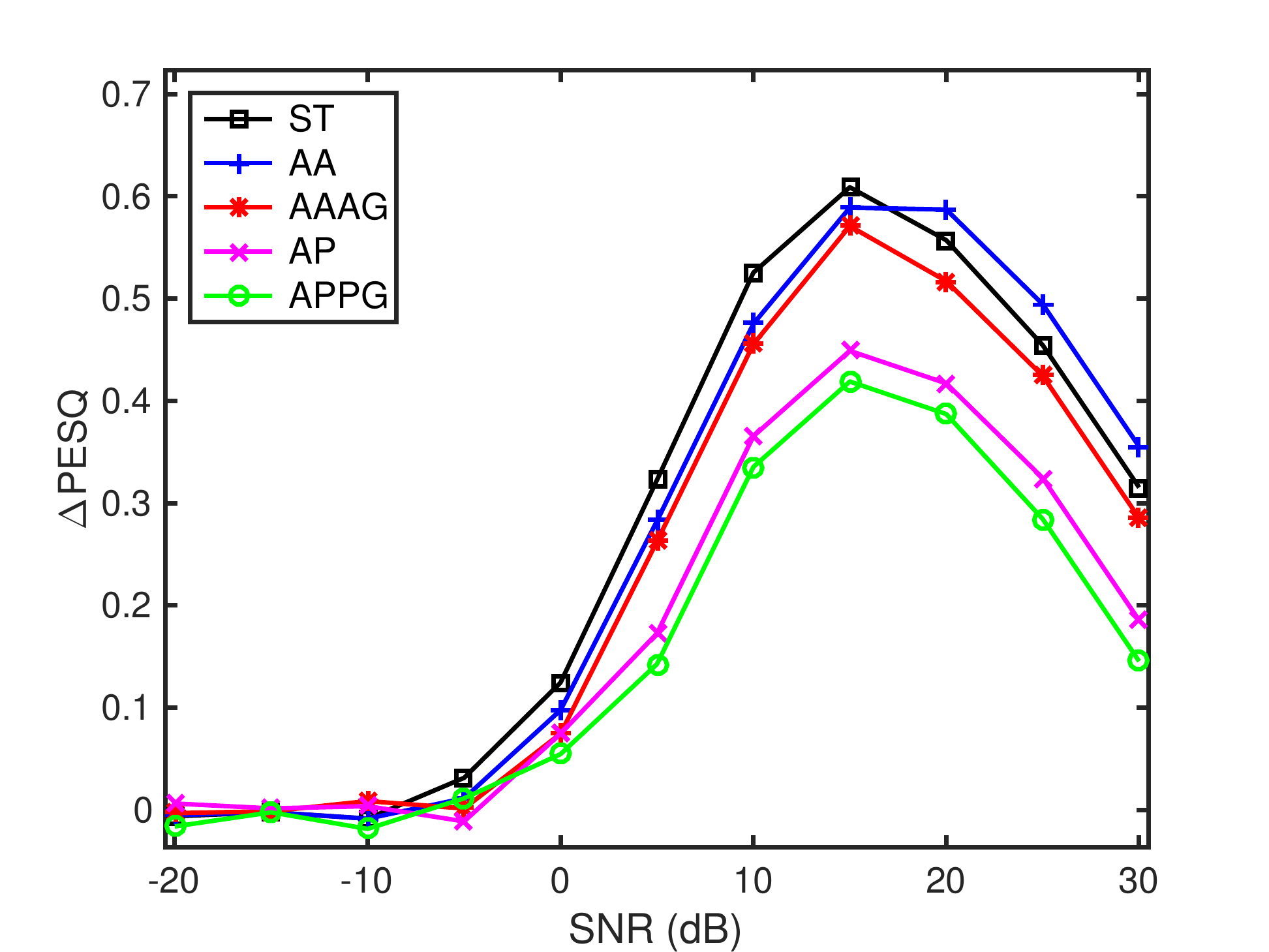}

\caption{Plot of the mean $\Delta$PESQ for speech with additive babble noise. }
\end{minipage}\hfill{}%
\begin{minipage}[t]{0.48\columnwidth}%
\centering \includegraphics[width=0.96\columnwidth]{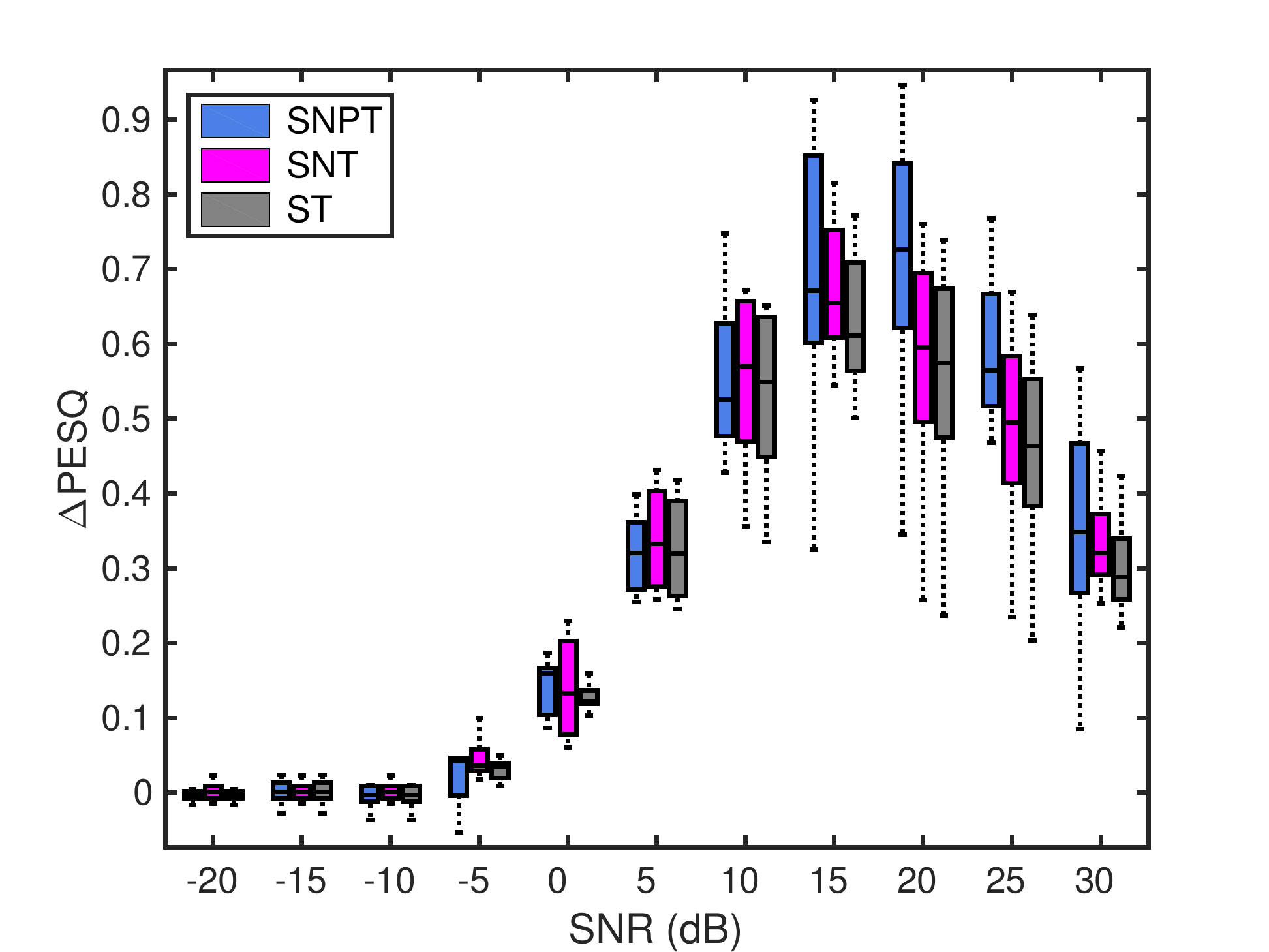}\caption{Boxplot of the $\Delta$PESQ for speech with babble noise versus SNR
showing the effect of tracking different quantities in the KF. }
\end{minipage}
\end{figure}

\subsection{Choice of signal model }

We now compare ST, which is presented in Sec. IV.A, with the alternative
KF updates presented in Sec. IV.B that use a simpler signal model
and assume $\alpha=0$ or $\alpha=1$. Both the ST algorithm and the
main enhancement algorithm SNPT, which is presented in Sec. III, use
the phase factor distribution. 

We compare Kalman filtering algorithms that use different signal models.
In Figs. 13,14, we compare ST with AP, AA, APPG and AAAG that are
presented in Sec. IV.B. We see that the algorithms that use $\alpha \in [-1,1]$
(i.e. ST) or assume $\alpha=1$ (i.e. AA and AAAG) outperform the
algorithms that assume $\alpha=0$ (i.e. AP and APPG) by a large margin,
in terms of PESQ. In addition, in Figs. 13,14, we also observe that
using $\alpha \in [-1,1]$ is more beneficial than assuming $\alpha=1$
for low SNR levels and a lot more beneficial than assuming $\alpha=0$.
For most SNRs, we have: $\text{AA}\geq\text{AAAG}\geq\text{AP}\geq\text{APPG}$. 

Therefore, based on Figs. 13,14, utilising $\alpha \in [-1,1]$ and
$p(\alpha)$ in the signal model, and thus using that speech and noise
are additive in the complex STFT domain, is beneficial. 

\subsection{Choice of tracked quantities }

The next step is to compare the more complex versions of the Kalman
filtering algorithm, SNT and SNPT, with ST. 

We now examine the $\Delta$PESQ of the SNT algorithm and we compare
SNT with ST. We use babble noise in Fig. 15 and we observe that SNT
is better than ST for $0 \leq \text{SNR} \leq 30$ dB. With NT, we
track the noise log-spectrum and the correlation between the speech
and noise log-spectra. The nonlinear KF update step presented in Sec.
III.C and Sec. III.D estimates the correlation between the speech
and noise log-spectra. With NT, we aim to approach the algorithm's
maximum performance that is defined by the oracle knowledge of the
actual noise power in each time-frequency cell. If we assume oracle
knowledge of the actual noise power in each time-frequency cell, the
median $\Delta$PESQ at the SNR of $15$ dB goes to $\Delta\text{PESQ}=1.08$
in Fig. 15 for babble noise. 

We now concentrate on SNPT. Figure 15 is a boxplot that depicts the
median and the interquartile range of the $\Delta$PESQ scores of
SNPT, SNT and ST for babble noise. Figure 15 shows that SNPT is better
than SNT for $10 \leq \text{SNR} \leq 30$ dB. We observe that speech
phase KF tracking is effective at high SNRs. Moreover, Fig. 15 shows
that the SNPT, SNT and ST algorithms do not have comparable $\Delta$PESQ
variability. At SNRs from $15$ dB to $30$ dB, the $\Delta$PESQ
variability of SNPT is larger than the $\Delta$PESQ variability of
SNT and ST. 

For F16 noise, SNPT is better than SNT for $5 \leq \text{SNR} \leq 30$
dB. Speech phase tracking is effective at high SNRs. 

As in \cite{a126}, to assess the robustness to noise type, we evaluate
the algorithms using the four different noise types in Fig. 6 with
the average SNR for each noise type chosen to give a mean PESQ score
of $2.0$ for the noisy speech. Combining different noise types into
a single graph, Fig. 16 examines the average of the $\Delta$PESQ
scores for the noise types of car, babble, F16 and white. Based on
Fig. 16 and on the $\Delta$PESQ results in Sec. VI, the phase-aware
ST block contributes the most to the final $\Delta$PESQ of SNPT.
The next most beneficial block is NT. Moreover, PT contributes further
to enhancement, mainly for the SNR range of $5 < \text{SNR} \leq 30$
dB. 

Figure 17 shows the average of the difference in $\Delta$PESQ between
competing algorithms and SNPT for the noise types of car, babble,
F16 and white. Figure 17 uses the noise types in Fig. 6 with the average
SNR for each noise type chosen to give a mean PESQ score of $2.0$
for the noisy speech. According to Fig. 17, the proposed SNPT algorithm
results in an improvement in $\Delta$PESQ and thus in speech quality. 

\begin{figure}[t]
\begin{minipage}[t]{0.48\columnwidth}%
\centering \includegraphics[width=0.99\columnwidth]{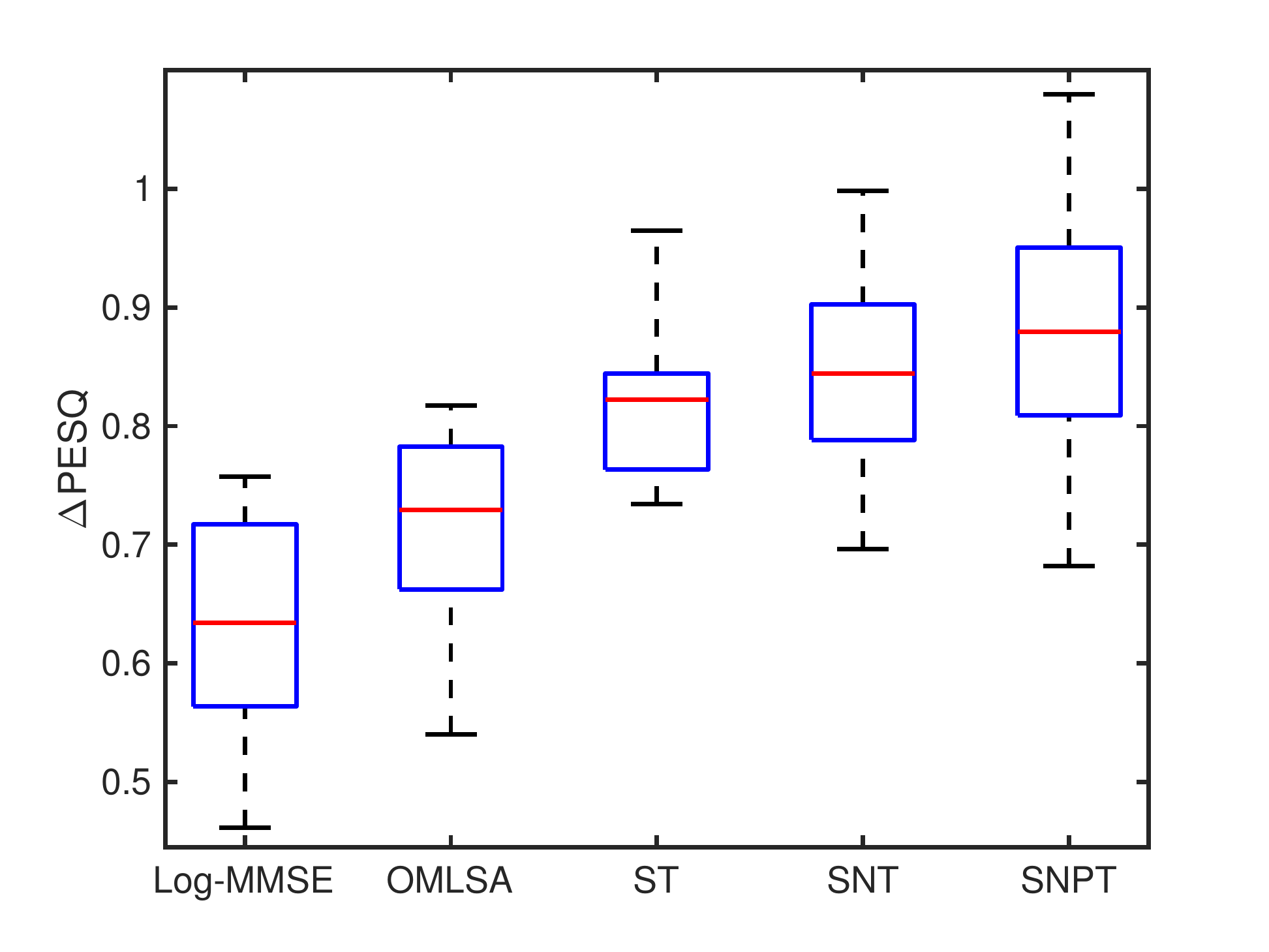}

\caption{Boxplot of the average of the $\Delta$PESQ scores for the noise types
of car, babble, F16 and white. }
\end{minipage}\hfill{}%
\begin{minipage}[t]{0.48\columnwidth}%
\centering \includegraphics[width=0.99\columnwidth]{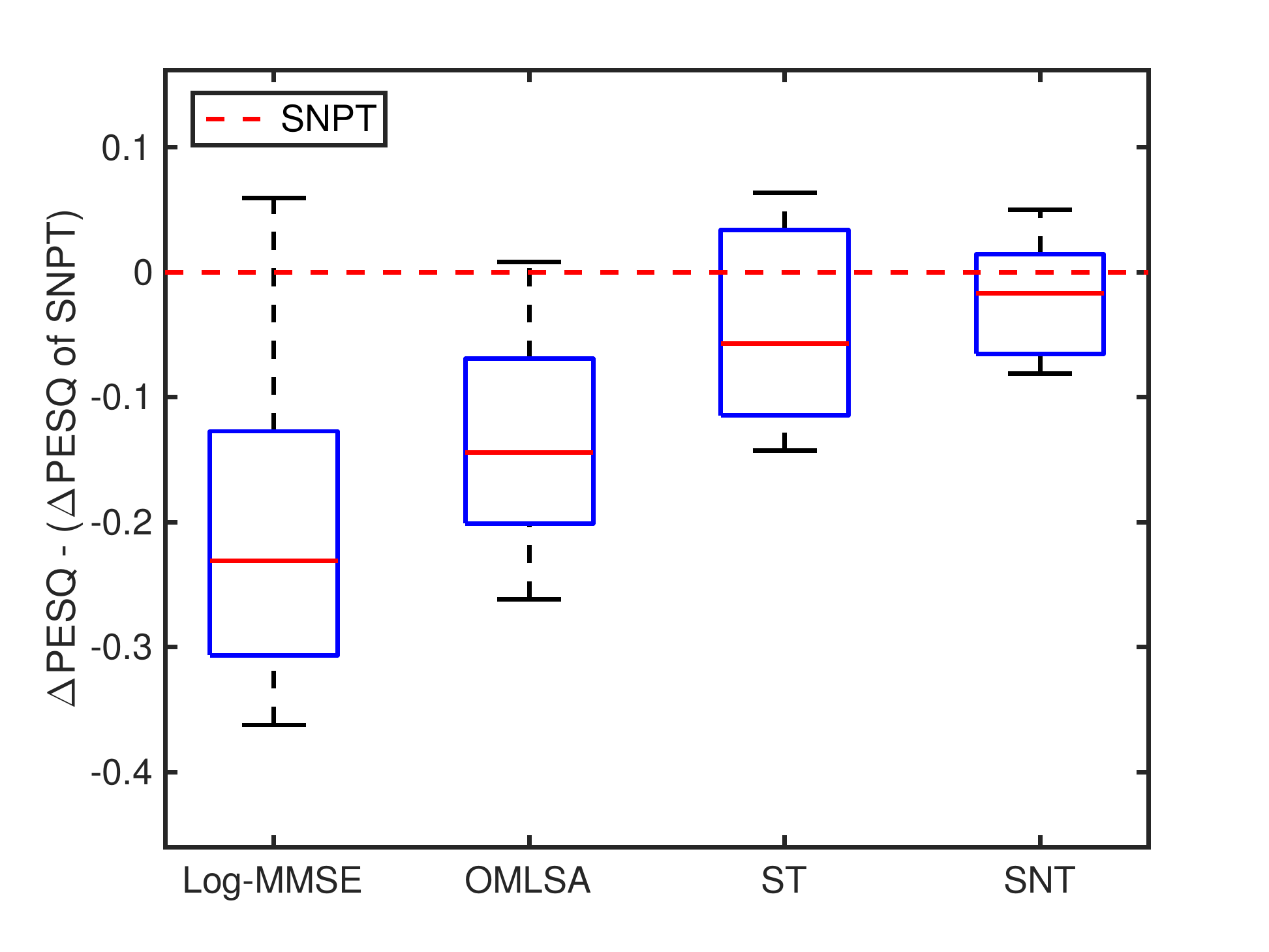}

\caption{Boxplot of the average of the difference in $\Delta$PESQ between
competing algorithms and SNPT for car, babble, F16 and white noises. }
\end{minipage}
\end{figure}

\ctable[caption = \textrm{\normalfont The $\Delta$SIG, $\Delta$BAK, $\Delta$OVRL at different SNRs (dB) for babble noise}.,   
label = tab:fpsaaaa6a324234asafda55aasf22232ahhaaaaaaaiiuuuaaagaaa0a0a0aaa0aaaaaooiiiaiaiaaaaa,   
pos = hp,   
doinside=\hspace*{0pt}, width=.49\textwidth ]{p{2.1cm} p{1.8cm} p{1.8cm} p{1.8cm}}{}{ 
\FL & \textbf{SNR = 5} & \textbf{SNR = 10} & \textbf{SNR = 15} 
\ML ST 
\NN $\Delta$SIG & 0.19 & 0.40 & 0.48 
\NN $\Delta$BAK & 0.33 & 0.45 & 0.51 
\NN $\Delta$OVRL & 0.26 & 0.48 & 0.59 
\ML SNPT 
\NN $\Delta$OVRL & 0.23 & 0.49 & 0.61 
\ML Log-MMSE 
\NN $\Delta$OVRL & 0.06 & 0.22 & 0.34 
\ML OMLSA 
\NN $\Delta$OVRL & -0.11 & 0.19 & 0.43 
\LL } 

\subsection{Alternative quality metrics }

Speech quality metrics other than PESQ are now utilised. We use the
SIG, BAK and OVRL metrics \cite{a205} \cite{a184}, which are on
a scale of $1$ to $5$ where $5$ indicates excellent quality. Table
II shows the SIG, BAK and OVRL of the ST, SNPT, OMLSA and Log-MMSE
algorithms for babble noise. 

Based on Table II, the proposed ST and SNPT algorithms improve speech
quality in terms of SIG, BAK and OVRL. 

\section{Conclusion }

\label{sec:conclusions-1-3} In this paper, we present a phase-aware
single-channel enhancement algorithm that is based on tracking the
time evolution of the speech log-spectrum and phase. We create a KF
prediction step that models the inter-frame relations of the speech log-spectrum,
the speech phase and the noise log-spectrum. In addition, we create
and implement a phase-sensitive nonlinear KF update step that models the
nonlinear relations between the speech log-spectrum, the speech phase
and the noise log-spectrum. The nonlinear KF update step guides the model towards a good KF posterior, thus suppressing noise, using transformations of random variables and using $(s,n,\phi,\psi) \Rightarrow (u,y,\gamma,\theta)$ in Sec. III.D. We utilise both the phase difference between
noise and speech, $\gamma$, and the phase difference between noisy
speech and speech, $\delta$. The time-varying KF enhances
the temporal characteristics of the trajectories of the speech log-spectrum
that are distorted by noise. With modulation-domain Kalman filtering, temporal contraints are imposed on the log-spectrum. Based on the evaluation in Sec. VI, the presented Kalman filtering algorithm improves speech quality; instrumental
measures predict a speech quality increase over a range of SNRs for
various noise types. 


\selectlanguage{british}%
\bibliographystyle{IEEEtran}
\addcontentsline{toc}{section}{\refname}\bibliography{abntex2-options,sapref}

\end{document}